\documentclass[aps,prb,showpacs,superscriptaddress,twocolumn,floatfix]{revtex4}
\usepackage{graphicx}
\usepackage{bm}
\usepackage{bbm}
\usepackage{latexsym}
\usepackage{amssymb}
\usepackage{amsmath}

\begin{document}
\title{Thermodynamics of Heisenberg ferromagnets with arbitrary spin 
in a magnetic field}
\author{I. Juh\'asz Junger}
\affiliation{Institut f\"{u}r Theoretische Physik,
Universit\"{a}t Leipzig, D-04109 Leipzig, Germany}
\author{D. Ihle}
\affiliation{Institut f\"{u}r Theoretische Physik,
Universit\"{a}t Leipzig, D-04109 Leipzig, Germany}
\author{L. Bogacz}
\affiliation{Institut f\"{u}r Theoretische Physik,
Universit\"{a}t Leipzig, D-04109 Leipzig, Germany}
\affiliation{Department of Information Technologies,
      Faculty of Physics, Astronomy and Applied Informatics,
      Jagellonian University, 30-059 Krak\'ow, Poland }
\author{W. Janke}
\affiliation{Institut f\"{u}r Theoretische Physik,
Universit\"{a}t Leipzig, D-04109 Leipzig, Germany}
\affiliation{Centre for Theoretical Sciences (NTZ), Universit\"at Leipzig, 
D-04105 Leipzig, Germany}

\date{\today}

\begin{abstract}
The thermodynamic properties (magnetization, magnetic susceptibility, 
transverse and longitudinal correlation lengths, specific heat) of 
one- and two-dimensional ferromagnets with 
arbitrary spin $S$ in a magnetic field are investigated by a second-order
Green-function theory. 
In addition, quantum Monte Carlo simulations for $S= 1/2$ and $S=1$ 
are performed using the stochastic series expansion method. A good 
agreement between the results of both approaches is found. The field
dependence of the position of the maximum in the temperature dependence of 
the susceptibility fits well to a power law at low fields and to a linear 
increase at high fields. The maximum height decreases according to a power  
law in the whole field region. The longitudinal correlation length may 
show an anomalous temperature dependence: a minimum followed by a maximum 
with increasing temperature. 
Considering the specific heat in one dimension and at low magnetic 
fields, two maxima in its temperature dependence for both the $S= 1/2$ 
and $S = 1$ ferromagnets are found.  For $ S>1$ only one maximum occurs,  
as in the two-dimensional ferromagnets. Relating the theory to experiments on 
the $S= 1/2$ quasi-one-dimensional copper salt TMCuC [(CH$_3$)$_4$NCuCl$_3$],    
a fit to  
the magnetization as a function of the magnetic field yields the value of the 
exchange energy which is used to make predictions for the occurrence of 
two maxima in the temperature dependence of the specific heat.
\end{abstract} 
\maketitle

\section{INTRODUCTION}\label{intro}
%\vspace*{2mm}
The study of low-dimensional quantum spin systems\cite{SRF04} 
is of growing interest and is motivated by the progress in the synthesis 
of new materials, where ferromagnetic compounds attract increasing attention. 
For example, besides the spin $S=1/2$ 
quasi-one-dimensional (1D) ferromagnetic systems, such as the copper salt 
TMCuC,\cite{LW79,DRS82}  the organic magnets p-NPNN\cite{TTN91,TKI92} and 
$\beta$-BBDTA$\cdot$GaBr$_4$,\cite{SGM06} and the CuCl$_2$-sulfoxide 
complexes,\cite{SLW79} recently the $S=1/2$ 
quasi-2D ferromagnet Cs$_2$AgF$_4$, which has a structure similar to the 
high-$T_c$ parent compound La$_2$CuO$_4$, was studied\cite{MTT05} and 
found to be magnetically reminescent of K$_2$CuF$_4$.\cite{LPS87} In 
ferromagnetic systems with $S \geqslant 1$ mainly the effects of single-ion
spin anisotropies were investigated, such as in the quasi-1D $S=1$ easy-plane 
ferromagnet 
CsNiF$_3$\cite{KV87} and in 2D easy-axis Heisenberg models in a magnetic
field\cite{FJK00,HFK02,SKN05} for describing the spin reorientation transition
in thin ferromagnetic films (see Ref.~\onlinecite{SFM01}
and references therein). \\[-4.5mm]

The 2D anisotropic $S \geqslant 1$ Heisenberg ferromagnets in a magnetic field
were investigated by Green-function methods,\cite{FJK00,FKS02,SKN05} where
the exchange term was treated in the random-phase approximation 
(RPA),\cite{Tjab67} and by quantum Monte Carlo (QMC) simulations.\cite{HFK02}
In a previous paper\cite{JIR04} we have developed a second-order Green-function
theory of 1D and 2D $S=1/2$ ferromagnets in a magnetic field which goes one 
step beyond the RPA and provides a rather good description of
magnetic short-range order (SRO) and of the thermodynamics. This can be seen 
from the comparison with the exact calculations by the Bethe-ansatz method 
for the quantum transfer matrix in the 1D model and with the exact 
diagonalizations on finite lattices. In particular, for the 
$S=1/2$ ferromagnetic chain two maxima in the temperature dependence of 
the specific heat at very low magnetic fields were found. On the contrary, 
the RPA was shown to fail in describing the SRO, reflected, e.g., in the
specific heat, whereas the magnetization and the magnetic susceptibility
are quite well reproduced. Recently, a similar Green-function approach for 
$S=1/2$ ferromagnets was presented\cite{APP07} which improves the 
theory of Ref.~\onlinecite{JIR04} concerning the agreement with exact methods. 
The results obtained for $S=1/2$ are stimulating to investigate 
ferromagnets with $S>1/2$ in a magnetic field, for which a second-order 
Green-function theory of SRO is not yet developed. Second-order Green-function 
approaches for ferromagnets with arbitrary spin exist in the case of 
zero magnetic field only,\cite{SSI94,JIR05} where in Ref.~\onlinecite{JIR05} 
ferromagnetic chains with an easy-axis single-ion anisotropy were studied.   

In this paper we extend both our previous theory for $S=1/2$ 
(Ref.~\onlinecite{JIR04}) to arbitrary spins and the theory of 
Ref.~\onlinecite{SSI94} for zero field to arbitrary fields. We start from 
the ferromagnetic Heisenberg model with arbitrary spin $S$,
\begin{equation}
H=-J \sum_{\langle ij \rangle} \bm{S}_{i} \bm{S}_{j} -
h \sum_{i} S_{i}^{z}
\label{ham}
\end{equation}
[$\langle ij \rangle$ denote nearest-neighbor (NN) bonds along a chain
or on a square lattice; throughout we set $J=1$] with $\bm{S}_{i}^{2}=S(S+1)$. 
We calculate thermodynamic properties 
(magnetization, magnetic susceptibility, correlation length, 
specific heat) at arbitrary temperatures and fields. 
For comparison, we perform QMC simulations of the $S=1/2$ and $S=1$ models 
on a chain up to $N=L=1024$ sites and on a square lattice up to 
$N=L \times L=64 \times 64$. 

The rest of the paper is organized as follows. In Sec.~\ref{gfth} 
the second-order Green-function 
theory for model (\ref{ham}) is developed, where the extensions of previous 
second-order Green-function approaches\cite{JIR04,SSI94,APP07} to arbitrary 
spins and fields imply novel technical aspects. Moreover, considering the case 
$S=1/2$ the theory is extended as compared with Refs.~\onlinecite{JIR04} and 
\onlinecite{APP07} by the introduction of two additional vertex parameters and, 
correspondingly, by taking into consideration two additional conditions for 
their determination. This extension is shown to have qualitative effects on the 
temperature dependence of the longitudinal correlation length 
(see Sec.~\ref{sec:res}.B). 
In Sec.~\ref{mcsim} the employed QMC method is briefly described. 
In Sec.~\ref{sec:res}  
the thermodynamic properties of the 1D and 2D ferromagnets are investigated 
as functions of temperature and field, also in comparison with RPA, 
and are related to experiments. 
Particular attention is paid to the calculation of the transverse and 
longitudinal correlation lengths which were not considered in 
Refs.~\onlinecite{JIR04} and \onlinecite{APP07}. Finally, a summary of our 
work is given in 
Sec.~\ref{summ}.

\section{SECOND-ORDER GREEN-FUNCTION THEORY}\label{gfth}
To determine the transverse and longitudinal spin correlation functions and
the thermodynamic quantities, we employ the equation of motion method for 
two-time retarded commutator Green functions.\cite{Tjab67} First we calculate 
the transverse spin correlation functions. Because we treat arbitrary spins in 
nonzero magnetic fields, so that we have $\langle S^z \rangle \neq 0$, we 
consider the Green functions 
$ \langle \langle S_{\bm{q}}^{+}; S_{-\bm{q}}^{(n)-} \rangle \rangle_{\omega} $ 
introduced by Tyablikov within the first-order theory, i.e. the RPA (see 
Appendix), where $S_{-\bm{q}}^{(n)-}$ is the Fourier transform of 
$S_{i}^{(n)-}=(S_{i}^{z})^{n} S_{i}^{-}$ with $n=0,1, ..., 2 S-1$, and 
the Green functions $ \langle \langle i \dot{S}_{\bm{q}}^{+}; S_{-\bm{q}}^{(n)-} 
\rangle \rangle_{\omega}$ which we calculate for the first time in the 
second-order theory. The equations of motion read
\begin{equation}  
\omega \langle \langle S_{\bm{q}}^{+}; S_{-\bm{q}}^{(n)-} 
\rangle \rangle_{\omega} = M^{(n)+-} + 
\langle \langle i \dot{S}_{\bm{q}}^{+}; S_{-\bm{q}}^{(n)-} 
\rangle \rangle_{\omega},
\label{eqmtn1}
\end{equation}
\begin{equation}  
\omega \langle \langle i \dot{S}_{\bm{q}}^{+}; S_{-\bm{q}}^{(n)-} 
\rangle \rangle_{\omega} = \tilde{M}_{\bm{q}}^{(n)+-} + 
\langle \langle - \ddot{S}_{\bm{q}}^{+}; S_{-\bm{q}}^{(n)-} 
\rangle \rangle_{\omega}.
\label{eqmtn2}
\end{equation}
The moments $M^{(n)+-}= \langle [S_{\bm{q}}^{+},S_{\bm{-q}}^{(n)-}] \rangle $ 
and $\tilde{M}_{\bm{q}}^{(n)+-}= \langle [i\dot{S}_{\bm{q}}^{+}, 
S_{\bm{-q}}^{(n)-}] \rangle $ are given by the exact expressions 
\begin{eqnarray}  
 M^{(n)+-} = 2 \langle (S^z)^{n+1} \rangle + %\nonumber \\
 (1-\delta_{n,0}) \sum_{k=1}^{n} 
\binom{n}{k} (-1)^{k} \times &&   \nonumber \\
 \{ S(S+1) \langle (S^z)^{n-k} \rangle \ 
+ \langle (S^z)^{n-k+1} \rangle- \langle (S^z)^{n-k+2} \rangle \}, && \;\;\;    
\label{mn}
\end{eqnarray}
\begin{eqnarray}  
 \tilde{M}_{\bm{q}}^{(n)+-} &=& 
z (1-\gamma_{\bm{q}}) \{ 2 C_{10}^{(n)zz}  
+ C_{10}^{(n)-+}  \nonumber \\ 
&+& (1-\delta_{n,0}) \sum_{k=1}^{n} \binom{n}{k} (-1)^{k}[S(S+1) 
\nonumber \\
& \times & (\delta_{k,n} \langle S^z \rangle  
+ (1-\delta_{k,n}) C_{10}^{(n-k-1)zz}) \nonumber \\
&+& C_{10}^{(n-k)zz}-C_{10}^{(n-k+1)zz}] \}+h M^{(n)+-} , \phantom{ml} 
\label{tmqn}
\end{eqnarray}
where 
$C_{nm}^{(n)-+} \equiv C_{\bm{R}}^{(n)-+}=
\langle S_{0}^{(n)-} S_{\bm{R}}^{+} \rangle$,
$C_{nm}^{(n)zz} \equiv C_{\bm{R}}^{(n)zz}=
\langle (S_{0}^{z})^{n+1} S_{\bm{R}}^{z} \rangle$, 
$\bm{R}= n \bm{e}_{x}+ m \bm{e}_{y}$, $\gamma_{\bm{q}}=\frac{2}{z}
\displaystyle{\sum_{i=1}^{z/2}} \cos q_{i}$, and $z$ is the coordination
number.  Deriving Eqs.~(\ref{mn}) and (\ref{tmqn}) the operator identity 
\begin{equation}
\bm{S}_{i}^{2}=S_{i}^{-} S_{i}^{+} + S_{i}^{z} + (S_{i}^{z})^2
\label{id}
\end{equation}
has been used. In Eq.~(\ref{eqmtn2}) the second derivative
$- \ddot{S}_{\bm{q}}^{+}$ is approximated as indicated in
Refs.~\onlinecite{SSI94, JIR05, JIR04,WI97,KY72,ST91,SRI04}. That means, in
$- \ddot{S}_{i}^{+}$ we decouple the products of operators along
NN sequences $\langle i,j,l \rangle$ as
\begin{equation}
S_{i}^{+} S_{j}^{+} S_{l}^{-}=\alpha_{1}^{+-} \langle S_{j}^{+} S_{l}^{-}
\rangle S_{i}^{+}+ \alpha_{2}^{+-} \langle S_{i}^{+} S_{l}^{-} \rangle
S_{j}^{+},
\label{entk1}
\end{equation}
where the vertex parameters $\alpha_{1}^{+-}$ and $\alpha_{2}^{+-}$ 
are attached to NN and further-distant 
correlation functions, respectively. The products of operators with two 
coinciding sites, appearing for $S \geqslant 1$, are decoupled 
as\cite{JIR05,SSI94}
\begin{equation}
S_{i}^{+} S_{j}^{-} S_{j}^{+}=\langle S_{j}^{-} S_{j}^{+}
\rangle S_{i}^{+}+ \lambda^{+-} \langle S_{i}^{+} S_{j}^{-} \rangle
S_{j}^{+},
\label{entk2}
\end{equation}
where the vertex parameter $\lambda^{+-}$ is introduced. We obtain 
\begin{equation}
-\ddot{S}_{\bm{q}}^{+}=[(\omega_{\bm{q}}^{+-})^2-h^2] S_{\bm{q}}^{+} +
2 h i \dot{S}_{\bm{q}}^{+}
\label{sqpp}
\end{equation}
with
\begin{equation}
(\omega_{\bm{q}}^{+-})^2 = \frac{z}{2}(1-\gamma_{\bf{q}})
\{ \Delta^{+-} + 2 z \alpha_{1}^{+-} C_{10} (1-\gamma_{\bf{q}}) \},
\label{omqp2}
\end{equation}
\begin{eqnarray}
 \Delta^{+-} &=&  S(S+1) + \langle (S^z)^2 \rangle   \nonumber \\
 & + & 2 \{ \lambda^{+-} - (z+1) \alpha_{1}^{+-} \} C_{10} \nonumber  \\
 &+ & 2 \alpha_{2}^{+-}\{ (z-2) C_{11} + C_{20} \},   \label{deltap} 
\end{eqnarray}
where $C_{nm}=\frac{1}{2} C_{nm}^{(0)-+} + C_{nm}^{(0)zz} $. In the special 
case $S=1/2$, in $-\ddot{S}_{i}^{+}$ products of spin operators with two 
coinciding sites do not appear which is equivalent to setting 
$\lambda^{+-}=0$. 
Finally, we get the Green functions 
\begin{equation}
\langle \langle S_{\bm{q}}^{+}; S_{-\bm{q}}^{(n)-}
\rangle \rangle_{\omega} = 
\sum_{i=1,2} \frac{A_{\bm{q}i}^{(n)}}{\omega-\omega_{\bm{q}i}},
\label{gf}
\end{equation}
\begin{equation}
\langle \langle i \dot{S}_{\bm{q}}^{+}; S_{-\bm{q}}^{(n)-}
\rangle \rangle_{\omega} = 
\sum_{i=1,2} \frac{\omega_{\bm{q}i} A_{\bm{q}i}^{(n)}}{\omega-\omega_{\bm{q}i}},
\label{hgf}
\end{equation}
where
\begin{equation}
\omega_{\bm{q}1,2}=h \pm \omega_{\bm{q}}^{+-},
\label{omq12}
\end{equation}
\begin{equation}
A_{\bm{q}1,2}^{(n)}=\frac{1}{2} M^{(n)+-} \pm \frac{1}{2 \omega_{\bm{q}}^{+-}} 
(\tilde{M}_{\bm{q}}^{(n)+-}- h M^{(n)+-})
\label{aq12}
\end{equation}
with the moments given by Eqs.~(\ref{mn}) and (\ref{tmqn}). The transverse 
dynamic spin susceptibility $\chi_{\bm{q}}^{+-} (\omega) = 
- \langle \langle S_{\bm{q}}^{+} ; S_{-\bm{q}}^{-} \rangle \rangle_{\omega}$ 
is given by Eq.~(\ref{gf}) for $n=0$.

Because we consider nonzero magnetic fields within the second-order theory, the  
behavior of the Green functions
(\ref{gf}) with the poles (\ref{omq12}) exhibits, for arbitrary 
spin, a peculiar aspect. Considering the static Green functions 
$ \langle \langle S_{\bm{q}}^{+} ; S_{-\bm{q}}^{(n)-} 
\rangle \rangle_{\omega=0}$, 
in particular the static spin susceptibility 
$\chi_{\bm{q}}^{+-} \equiv \chi_{\bm{q}}^{+-} (\omega=0)$, a divergency 
signaling a phase transition could appear if 
$\omega_{\bm{q}2}=0$, i.e., $\omega_{\bm{q}}^{+-}=h$. According to 
Eq.~(\ref{omqp2}) the corresponding $\bm{q}$ values are given by 
\begin{eqnarray}
 1-\gamma_{\bm{q}} &=& g_{0} \equiv (4 z \alpha_{1}^{+-} C_{10})^{-1} 
 \times  \nonumber \\ 
&& \{
[(\Delta^{+-})^2 + 16 \alpha_{1}^{+-} C_{10} h^2]^{1/2}- \Delta^{+-}
\}. \phantom{mm} \label{g0} 
\end{eqnarray}
This equation may be fulfilled in region I of the $h-T$ plane defined 
by $h<h_0(T)$, where $h_0(T)$ is determined by 
Eq.~(\ref{g0}) with $g_0=2$ which is realized at the corner of the 
Brillouin zone with $\gamma_{\bm{q}}=-1$. In region II, 
$h>h_0(T)$, we have $h>\omega_{\bm{q}}^{+-}$ for all $\bm{q}$. 
For nonzero fields the Heisenberg ferromagnet described by Eq.~(\ref{ham}) 
has no phase transition. This means, $\chi_{\bm{q}}^{+-}$ has to be finite 
at all $\bm{q}$. We require this regularity to hold also for the static Green 
functions with $n=1,..., 2S-1$. That is, in region I we require 
$A_{\bm{q}2}^{(n)}=0$ with $\bm{q}$ given by Eq.~(\ref{g0}). This results 
in the regularity conditions, 
\begin{eqnarray}
&& h M^{(n)+-} = z \{2 C_{10}^{(n)zz} + C_{10}^{(n)-+} +  \nonumber \\
&& (1-\delta_{n,0}) \sum_{k=1}^{n} \binom{n}{k} (-1)^{k} 
[S(S+1) (\delta_{k,n} \langle S^z \rangle + \nonumber \\ 
&& (1-\delta_{k,n}) C_{10}^{(n-k-1)zz}) 
  + C_{10}^{(n-k)zz}-C_{10}^{(n-k+1)zz}] 
\} g_0. \phantom{m}  \nonumber \\
\label{rb1}
\end{eqnarray}
Note that Eq.~(\ref{rb1}) for $S=1/2$ agrees with the condition given in 
Ref.~\onlinecite{APP07} which is obtained from an analyticity argument and 
is written as an expression for $\langle S^z \rangle$.
In the limit $T\to \infty$, the field $h_0$ separating the regions I and II
may be easily obtained. For $T \to \infty$ we have
spin rotational symmetry so that
$\langle (S^z)^2 \rangle =\frac{1}{2} C_{00}^{(0)-+}$. By Eq.~(\ref{id}) with  
$\displaystyle{\lim_{T \to \infty}} \langle S^{z} \rangle =0$ we get 
$\langle (S^z)^2 \rangle=\frac{1}{3} S(S+1)$ resulting in 
$\Delta^{+-}=\frac{4}{3}S(S+1)$ and 
$(\omega_{\bm{q}}^{+-})^2=\frac{z}{2} \Delta^{+-}
(1-\gamma_{\bm{q}})$. From $\omega_{\bm{q}}^{+-}=h$ 
and $g_0=2$ we get $\displaystyle{\lim_{T\to\infty}} h_0 (T)
= 2 \sqrt{z S(S+1)/3}$. 
Following Ref.~\onlinecite{APP07}, we assume the conditions (\ref{rb1}) to 
be valid also in region II. This guarantees the continuity of all quantities 
at the boundary $h_0(T)$.

From the Green functions (\ref{gf}) and (\ref{hgf}) the  
transverse correlators $C_{\bm{R}}^{(n)-+} = (1/N)
\displaystyle{\sum_{\bm{q}}} C_{\bm{q}}^{(n)-+} \text{e}^{i
\bm{q R}}$ and $\tilde{C}_{\bm{R}}^{(n)-+} = (1/N)
\displaystyle{\sum_{\bm{q}}} \tilde{C}_{\bm{q}}^{(n)-+} \text{e}^{i
\bm{q R}}$ with the structure factors $C_{\bm{q}}^{(n)-+}=
\langle S_{-\bm{q}}^{(n)-} S_{\bm{q}}^{+} \rangle$ 
and $\tilde{C}_{\bm{q}}^{(n)-+}=
\langle S_{-\bm{q}}^{(n)-} i \dot{S}_{\bm{q}}^{+} \rangle$
are calculated by the spectral theorem,
\parbox{7cm}
{\begin{eqnarray*}
C_{\bm{q}}^{(n)-+}&=&\sum_{i=1,2} A_{\bm{q}i}^{(n)} n(\omega_{\bm{q}i}), \\
\tilde{C}_{\bm{q}}^{(n)-+}&=&\sum_{i=1,2} \omega_{\bm{q}i} A_{\bm{q}i}^{(n)}
n(\omega_{\bm{q}i}),
%\label{cqp}
\end{eqnarray*}} \hfill
\parbox{1cm}{
\begin{eqnarray}
\label{cqp}
\end{eqnarray}}
where $n(\omega)=(\text{e}^{\beta \omega}-1)^{-1}$ and $\beta =1/T$.

Now we derive some useful sum rules. Using $\langle S_{i}^{(n)-} S_{i}^{+} 
\rangle = \langle (S_{i}^{z})^{n} S_{i}^{-} S_{i}^{+} \rangle$ obtained 
from Eq.~(\ref{id}) multiplied by $(S_{i}^{z})^{n}$ ($n=0,1, ... , 2S-1$) and 
Eq.~(\ref{cqp}) we get the relation
\begin{eqnarray}
&S(S+1) \langle (S^z)^{n} \rangle - \langle (S^z)^{n+1} \rangle 
- \langle (S^z)^{n+2} \rangle & \nonumber \\
& \displaystyle{= \frac{1}{N} \sum_{\bm{q}} \sum_{i=1,2} 
A_{\bm{q}i}^{(n)} n (\omega_{\bm{q}i})}. & 
\label{sr}
\end{eqnarray}
By the identity $\displaystyle{ \prod_{m=-S}^{S} (S_{i}^{z} -m)=0 }$ one 
can express 
$(S_{i}^{z})^{2S+1} $ appearing in Eq.~(\ref{sr}) for $n=2S-1$ 
in terms of lower powers of $S_{i}^{z}$ (Refs.~\onlinecite{Tjab67} and 
\onlinecite{JA00}),
\begin{equation}
(S_{i}^{z})^{2S+1}= \sum_{k=0}^{2S} \alpha_{k}^{(S)} (S_{i}^z)^{k},
\label{szpower}
\end{equation}
where the coefficients $\alpha_{k}^{(S)}$ are given in Ref.~\onlinecite{JA00}. 
From the system of the $2S$ equations (\ref{sr}) we can  
determine the magnetization $m=-2 \mu_{B} \langle S^z \rangle $.

Similarly, in the second-order theory higher-derivative sum rules may be derived 
which, for nonzero fields, provide $2S$ additional equations for determining the 
vertex parameters and some longitudinal correlators (see below). Multiplying 
$S_{i}^{(n)-}$ by $i \dot{S}_{i}^{+}= \displaystyle{\sum_{j (n.n.i)}} 
(S_{j}^{z} S_{i}^{+}- S_{j}^{+} S_{i}^{z}) + h S_{i}^{+}$ 
and using Eqs.~(\ref{hgf}), (\ref{omq12}), (\ref{cqp}) and (\ref{id}) 
we obtain 
\begin{eqnarray}
&z \{S(S+1) [\delta_{n,0} \langle S^z \rangle 
+ (1- \delta_{n,0}) C_{10}^{(n-1)zz}]  & \nonumber \\
& - C_{10}^{(n)zz} - C_{10}^{(n+1)zz} 
- C_{10}^{(n)-+} - C_{10}^{(n+1)-+} \}  & \nonumber  \\
& \displaystyle {= -\frac{1}{N} \sum_{\bm{q}} \sum_{i=1,2}
(-1)^{i} \omega_{\bm{q}i}^{+-} A_{\bm{q}i}^{(n)} n(\omega_{\bm{q}i})}. & 
\label{hsr}
\end{eqnarray}
The correlator $C_{10}^{(n+1)zz}$ for $n=2S-1$ may be expressed in terms of 
$\langle S^z \rangle$ and $C_{10}^{(n)zz}$ with $n \leqslant 2S-1$ by 
Eq.~(\ref{szpower}). Equally, $C_{10}^{(2S)-+}$ can be written in terms of 
$C_{10}^{(n)-+}$ ($n \leqslant 2S-1$) by the identity\cite{JA00} 
\begin{equation}
S_{i}^{-} (S_{i}^{z})^{2S} = S_{i}^{-} \sum_{k=0}^{2S-1} \delta_{k}^{(S,1)} 
(S_{i}^{z})^{k},
\label{jens}
\end{equation}
where the coefficients $\delta_{k}^{(S,1)}$ are given in Ref.~\onlinecite{JA00}.
The product $(S_{i}^{z})^{2S} S_{i}^{-}$ appearing in $C_{10}^{(2S)-+}$ 
can be deduced from Eq.~(\ref{jens})  
by the commutation relations for spin operators.
The sum rule (\ref{hsr}) for $n=0$ also follows from the exact
representation of the internal energy per site,
$u=\langle H \rangle/N =-\frac{z}{2} (C_{10}^{(0)-+}+C_{10}^{(0)zz})
-h \langle S^{z} \rangle$, in terms of
$\langle \langle S_{\bm{q}}^{+} ; S_{-\bm{q}}^{-}
\rangle \rangle_{\omega}$ which can be derived similarly as in 
Ref.~\onlinecite{EG79} for $S=1/2$,
\begin{eqnarray}
u&=& - \frac{z}{2}[S(S+1) \langle S^z \rangle - C_{10}^{(1)zz}
- C_{10}^{(1)-+}] - h \langle S^z \rangle \quad \quad \nonumber \\
& - & \frac{1}{N}\sum_{\bm{q}} \int_{-\infty}^{+\infty}
\frac{d \omega}{2 \pi} (\omega-h) Im
\langle \langle S_{\bm{q}}^{+} ; S_{-\bm{q}}^{-}
\rangle \rangle_{\omega} n(\omega), \phantom{l}
\label{u}
\end{eqnarray}
if the result (\ref{gf}) for $\langle \langle S_{\bm{q}}^{+} ; 
S_{-\bm{q}}^{-} \rangle \rangle_{\omega}$ ($n=0$) is inserted into 
Eq.~(\ref{u}).

To calculate the longitudinal spin correlation functions $C_{\bm{R}}^{(0)zz}$ 
from the Green function 
$\langle\langle S_{\bm{q}}^{z};S_{-\bm{q}}^{z} \rangle \rangle_{\omega} = 
- \chi_{\bm{q}}^{zz} (\omega)$, where $\chi_{\bm{q}}^{zz} (\omega)$ is the 
longitudinal dynamic spin susceptibility, we start from the equations of 
motion analogous to 
Eqs.~(\ref{eqmtn1}) and (\ref{eqmtn2}) and perform a second-order decoupling 
which is equivalent to the projection method with the basis 
($S_{\bm{q}}^{z} , i \dot{S}_{\bm{q}}^{z}$) neglecting the self-energy, as 
indicated in our previous papers.\cite{JIR04,JIR05} In 
$-\ddot{S}_{i}^{z}$ we adopt the decouplings\cite{SSI94,JIR04,JIR05} analogous 
to Eqs.~(\ref{entk1}) and (\ref{entk2}),
\begin{equation}
S_{i}^{z} S_{j}^{+} S_{l}^{-}=\alpha_{1}^{zz} \langle S_{j}^{+} S_{l}^{-}
\rangle S_{i}^{z},
\label{zentk1}
\end{equation}
\begin{equation}
S_{i}^{+} S_{j}^{z} S_{l}^{-}=\alpha_{2}^{zz} \langle S_{i}^{+} S_{l}^{-}
\rangle S_{j}^{z},
\label{zentk1a}
\end{equation}
where $\langle i, j, l \rangle $ form NN sequences, and
\begin{equation}
S_{i}^{-} S_{j}^{z} S_{j}^{+}= \lambda^{zz} \langle S_{i}^{-}
S_{j}^{+} \rangle S_{j}^{z}.
\label{zentk2}
\end{equation}
We obtain
\begin{equation}
\chi_{\bm{q}}^{zz} (\omega) = 
 - \frac{M_{\bm{q}}^{zz}}{\omega^{2}-(\omega_{\bm{q}}^{zz})^{2}}
\label{gfz}
\end{equation}
with $ M_{\bm{q}}^{zz}=\langle [i
\dot{S}_{\bm{q}}^{z},S_{-\bm{q}}^{z}]\rangle $ given by
\begin{equation}
M_{\bm{q}}^{zz}=z C_{10}^{(0)-+} (1- \gamma_{\bm{q}})
\label{mqz}
\end{equation}
and
\begin{equation}
(\omega_{\bm{q}}^{zz})^2=\frac{z}{2} (1- \gamma_{\bm{q}})
\{ \Delta^{zz}+ 2 z \alpha_{1}^{zz} C_{10}^{(0)-+} (1-
\gamma_{\bm{q}}) \},
\label{omqz}
\end{equation}
\begin{eqnarray}
\Delta^{zz}&=&2 \{ S(S+1)-\langle (S^z)^{2} \rangle  \nonumber \\ 
&+& [\lambda^{zz} - (z+1) \alpha_{1}^{zz}] C_{10}^{(0)-+} \nonumber  \\
&+& \alpha_{2}^{zz} [(z-2)C_{11}^{(0)-+} + C_{20}^{(0)-+}]
\}. 
\label{deltaz}
\end{eqnarray}
As for the transverse correlations [cf. Eq.~(\ref{deltap})], in the case 
$S=1/2$ we have $\lambda^{zz}=0$. The correlation functions
$C_{\bm{R}}^{(0)zz}$ are calculated from\cite{JIR04}
\begin{equation}
C_{\bm{R}}^{(0)zz}=\frac{1}{N} \sum_{\bm{q}( \not=0)}
C_{\bm{q}}^{zz} \text{e}^{i \bm{q R}}+ \langle S^z \rangle^2
\label{crz}
\end{equation}
with
\begin{equation}
C_{\bm{q}}^{zz}=
\frac{M_{\bm{q}}^{zz}}{2 \omega_{\bm{q}}^{zz}} [1+2 n(\omega_{\bm{q}}^{zz})].
\label{cqz}
\end{equation}
Let us consider the magnetic susceptibility $\chi=4 \mu_{B}^{2}\chi_{S}$ 
with $\chi_{S}=\partial \langle S^z \rangle / \partial h$, which we denote 
by isothermal susceptibility, and its relation to the Kubo susceptibility 
(\ref{gfz}). From the first and the second derivatives of the partition 
function with respect to $h$ we obtain the exact relation
\begin{equation}
\chi_{S}=\frac{1}{T} \sum_{\bm{R}} \bar{C}_{\bm{R}}^{(0)zz} = 
\frac{1}{T} \bar{C}_{\bm{q}=0}^{zz},
\label{susc}
\end{equation}
where $\bar{C}_{\bm{R}}^{(0)zz} = C_{\bm{R}}^{(0)zz} - \langle S^z \rangle^2$, 
and the Fourier transform reads 
$\bar{C}_{\bm{q}}^{zz} = C_{\bm{q}}^{zz} 
- N \langle S^z \rangle^{2} \delta_{q,0} $. By Eqs.~(\ref{gfz}) and 
(\ref{cqz}) the uniform static Kubo susceptibility 
$\chi_{0}^{zz} = \displaystyle{\lim_{q \to 0} \lim_{\omega \to 0}}
\chi_{\bm{q}}^{zz} (\omega)$ 
may be expressed as
$\chi_{0}^{zz} = \frac{1}{T} \displaystyle{\lim_{q \to 0}} C_{\bm{q}}^{zz} = 
\textstyle{\frac{1}{T}} \displaystyle {\lim_{q \to 0}} \bar{C}_{\bm{q}}^{zz} 
= \textstyle{\frac{1}{T}} \bar{C}_{\bm{q}=0}^{zz}$. 
That is, within our theory the isothermal and Kubo susceptibilities 
agree at arbitrary fields and temperatures. Using Eqs.~(\ref{gfz}) to 
(\ref{omqz}) we have
\begin{equation}
\frac{\partial \langle S^z \rangle}{\partial h} = 
\frac{2 C_{10}^{(0)-+}}{\Delta^{zz}}.
\label{susc_eq}
\end{equation}
The equality (\ref{susc_eq}) is an additional equation for determining 
the parameters of the theory. 

Considering the ground state, at $T=0$ we have the exact results
\begin{equation}
C_{\bm{R}}^{(n)-+}(0)=0, \; C_{\bm{R}}^{(n)zz}(0)=S^{2+n}, \;
\langle (S^z)^{n} \rangle(0)=S^{n}.
\label{crpm}
\end{equation}
The regularity conditions (\ref{rb1}) read as $g_0=h/zS$. From $g_0=2$ the 
field $h_0(0)$ is given by $h_0(0)=2 z S$. Taking $g_0$ from 
Eq.~(\ref{g0}) we get the equation 
$\Delta^{+-}=2 (1-\alpha_{1}^{+-}) h S$. This equation can be fulfilled only, 
if $\alpha_{1}^{+-}(0)=1$ and $\Delta^{+-} (0) =0$, because in the ground 
state of the ferromagnet at $h \neq 0$ all quantities do not depend on $h$. 
Taking $\Delta^{+-}$ from Eq.~(\ref{deltap}) we get the parameter relation 
$\lambda^{+-}(0)+(z-1) \alpha_{2}^{+-}(0)=z-1/2S$. 
For $S=1/2$ ($\lambda^{+-}=0$) we have $\alpha_{2}^{+-} (0) = 1$.
Concerning the
zero-temperature values of $\alpha_{1}^{zz}$ and $\Delta^{zz}$, they can
be determined only in the limit $T \to 0$, since Eqs.~(\ref{crz}) and 
(\ref{cqz}) for $C_{\bm{R}}^{(0)zz}$ contain $M_{\bm{q}}^{zz}$ with 
$\lim_{T \to 0} M_{\bm{q}}^{zz}=0$.

\begin{figure}[b]
\includegraphics*[scale=0.581]{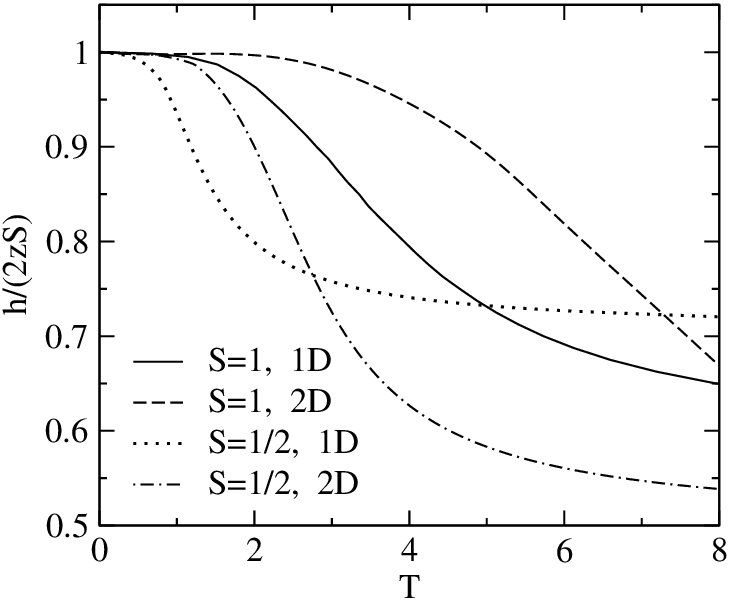}
\centering
\caption{Boundary $h_0(T)$ in the $h-T$ plane separating region I,
$h<h_0(T)$, where the equality $\omega_{\bm{q}}^{+-}=h$
[cf. Eq.~(\ref{omq12})] may be fulfilled, from region II, $h>h_0(T)$, where 
$h> \omega_{\bm{q}}^{+-}$ for all $\bm{q}$.}
\label{fig_h0}
\end{figure}

To evaluate the thermodynamic properties for arbitrary spin,  
the transverse correlators $C_{10}^{(n)-+}$, the 
longitudinal correlators ($\langle (S^z)^{n+1} \rangle$, $C_{10}^{(n)zz}$), 
and the parameters 
$\alpha_{1}^{\nu \mu}$ and $\Delta^{\nu \mu}$ ($\mu \nu=-+,\;zz $) 
have to be determined as solutions of a coupled system of 
self-consistency equations for arbitrary temperatures and fields. 
Note that for $S>1/2$ the parameters $\alpha_{2}^{\nu \mu}$ and 
$\lambda^{\nu \mu}$ have not to be calculated separately, because they only 
appear in the combination given by $\Delta^{\nu \mu}$.
The correlation functions $C_{10}^{(n)-+}$ are calculated from the Green 
functions according to Eqs.~(\ref{cqp}). 
To determine the $4(S+1)$ quantities $\langle (S^z)^{n+1} \rangle$ 
and $C_{10}^{(n)zz}$ with $n=0, ...,\; 2S-1$, $\alpha_{1}^{\nu \mu}$, 
and $\Delta^{\nu \mu}$, we have $6 S+3$ equations,  
namely the regularity conditions (\ref{rb1}), the sum rules 
(\ref{sr}) and (\ref{hsr}), Eqs.~(\ref{crz}) for $\langle (S^z)^2 \rangle $ 
and $C_{10}^{(0)zz}$, and the equality (\ref{susc_eq}). That is, for $S > 1/2$ 
we have $2 S-1$ more equations than quantities to be determined. 
To obtain a closed system of self-consistency equations for $S>1/2$, i.e. to 
reduce the number of equations (in addition to those for $C_{10}^{(n)-+}$) 
to $4 (S+1)$, we consider two choices. First we take 
into account the higher-derivative sum rule (\ref{hsr}) with $n=0$ only. 
As revealed by numerical evaluations, the specific heat of the 1D model 
strongly deviates from the QMC data for $S=1$, and for $S>1$ it 
even becomes negative 
at low fields and temperatures. Therefore, we adopt another choice, which 
yields a good agreement of all thermodynamic quantities with the QMC data 
for $S=1$ and which is used for $S \geqslant 1$ throughout the paper. 
Namely, we take into account the higher sum rules (\ref{hsr}) with $n=0$ 
and with $n=1$ instead of Eq.~(\ref{crz}) for $C_{10}^{(0)zz}$. To  
justify this choice within the theory itself, the correlator 
$C_{10}^{(0)zz}$ resulting from the closed system of equations is compared 
with $C_{10}^{(0)zz}$ calculated by Eq.~(\ref{crz}). For example, in the 
1D $S=1$ model at the fields $h=0.05$ and 0.1 the deviation is found to 
be less than 2\% at all temperatures except for the region 
$0.1 \lesssim T \lesssim 1$, where the maximal deviation is   
about 9\% for $T \simeq 0.3$ and 0.4, respectively. From the solution of 
the self-consistency equations in region I and from 
Eq.~(\ref{g0}) with $ g_{0}=2$ the boundary between regions 
I and II, $h_{0}(T)$, is determined. In Fig.~\ref{fig_h0}, $h_{0}(T)$ is 
plotted for $S=1/2$ and $S=1$. Note that in experiments 
realistic values of temperature and field lie in region I. Therefore, 
below nearly all results are presented in this region, and only some results 
for high enough temperatures and fields in region II are shown in 
Fig.~\ref{fig_chi_max}.
%
% fig_1

Let us finally make some comments on the evaluation of the theory for different 
spin values.\\    
(i) $S=\frac{1}{2}$: Using the identities $(S_{i}^{z})^2 =1/4$ and 
$S_i^z S_i^- = - \frac{1}{2}S_i^-$ (cf. Eq.~(\ref{jens})) the sum rules 
(\ref{sr}) and (\ref{hsr}) for $n=0$ simplify, where the higher sum rule 
(\ref{hsr}) reduces to 
\begin{equation}
z \left( \frac{1}{2} \langle S^z\rangle -C_{10} \right) =
-\frac{1}{N}\sum_{\bm{q},i} (-1)^{i} \omega_{\bm{q}i}^{+-}
A_{\bm{q}i}^{(0)} n(\omega_{\bm{q}i}).
\label{hsr_05}
\end{equation}
Note that this sum rule may be also obtained from the exact representation
(\ref{u}) of the internal energy which in the case $S=1/2$ becomes
(cf. Ref.~\onlinecite{JIR04})
\begin{eqnarray}
u &=&-\frac{z}{8}-\frac{h}{2}  \nonumber\\
&-& \frac{1}{N}\sum_{\bm{q}}
\int_{-\infty}^{+\infty} \frac{d \omega}{2 \pi}
(\varepsilon_{\bm{q}}+\omega) Im \langle\langle
S_{\bm{q}}^{+} ; S_{-\bm{q}}^{-} \rangle\rangle_{\omega} n(\omega) 
\phantom{mml} 
\label{inen}
\end{eqnarray}
with $\varepsilon_{\bm{q}}=z (1-\gamma_{\bm{q}})/2+h$, if $\langle\langle
S_{\bm{q}}^{+} ; S_{-\bm{q}}^{-} \rangle\rangle_{\omega}$ 
given by Eq.~(\ref{gf}) for $n=0$ is inserted into Eq.~(\ref{inen}). 
The spectra $\omega_{\bm{q}}^{+-}$ and 
$\omega_{\bm{q}}^{zz}$ are given by Eqs.~(\ref{omqp2}), (\ref{deltap}), 
(\ref{omqz}), and (\ref{deltaz}) with $\lambda^{+-}=0$ 
and $\lambda^{zz}=0$. We have to solve a closed system of coupled 
self-consistency equations for the seven quantities 
$\langle S^z \rangle$, $C_{10}^{(0) \mu \nu}$, $\alpha_{1}^{\nu \mu}$, 
and $\Delta^{\nu \mu}$ (or $\alpha_{2}^{\nu \mu}$). 
Note that in previous approaches\cite{JIR04, APP07} the simplified choice 
$\alpha_{2}^{\nu \mu}=\alpha_{1}^{\nu \mu}$ is taken disregarding the 
equality (\ref{susc_eq}) and not using either the condition (\ref{rb1}) 
(Ref.~\onlinecite{JIR04}) or the higher sum rule (\ref{hsr_05}) 
(Ref.~\onlinecite{APP07}).
\\
(ii) $S \geqslant 1$: Let us specify the identities (\ref{szpower}) and 
(\ref{jens}) which are used to reduce the sum rules (\ref{sr}) and 
(\ref{hsr}) for $n=2 S-1$, respectively, for $S=1$ and $S=3/2$. For $S=1$ 
we have $(S_{i}^{z})^{3}=S_{i}^{z}$ and 
$(S_{i}^{z})^{2} S_{i}^{-}= - S_{i}^{z} S_{i}^{-}$, and for $S=3/2$ we get 
$(S_i^z)^4=\frac{5}{2}(S_i^z)^2 - \frac{9}{16}$ 
and  
$(S_i^z)^3 S_{i}^{-}= -\frac{3}{2} (S_i^z)^2 S_{i}^{-} + 
  \frac{1}{4} S_i^z S_{i}^{-}+ \frac{3}{8} S_{i}^{-}$.
For $S=1$ a closed system of coupled self-consistency 
equations for the ten quantities $\langle S^{z} \rangle$, 
$\langle (S^{z})^2 \rangle$, $C_{10}^{(0) \mu \nu}$,  
$C_{10}^{(1) \mu \nu}$, $\alpha_{1}^{\nu \mu}$, and $\Delta^{\nu \mu}$ 
has to be solved.                               

In the case $h=0$ we have $\langle S^z \rangle =0 $, and the correlators for  
$n=0$ only are needed. The spin-rotation symmetry, implying $C_{\bm{R}}^{(0)-+}=
2 C_{\bm{R}}^{(0)zz} = C_{\bm{R}}$, is preserved by the second-order theory 
with $\alpha_{1,2}^{+-} = \alpha_{1,2}^{zz} \equiv \alpha_{1,2} $ and 
$ \lambda^{+-} = \lambda^{zz} \equiv \lambda $. 
Using $ \langle (S^z)^2 \rangle = \frac{1}{3} S(S+1) $ following from 
Eq.~(\ref{id}), the Eqs.~(\ref{omqp2}), (\ref{deltap}), (\ref{omqz}), and 
(\ref{deltaz}) yield the spectrum $ \omega_{\bm{q}}^{+-} = \omega_{\bm{q}}^{zz}
\equiv \omega_{\bm{q}} $ given by
\begin{equation}
\omega_{\bm{q}}^{2} = \frac{z}{2} (1-\gamma_{\bm{q}}) \{ \Delta 
+ 2 z \alpha_{1} C_{10} (1-\gamma_{\bm{q}}) \}
\label{omq2_05}
\end{equation}
with
\begin{eqnarray}
\Delta&=&\frac{4}{3} S (S+1)+ 2 \{ \lambda - (z+1) \alpha_{1}\} C_{10} \nonumber \\ 
&+& 2 \alpha_{2} \{ (z-2) C_{11} + C_{20} \},
\label{delta_05}
\end{eqnarray}
which agrees with the result of Ref.~\onlinecite{SSI94}, if we put 
$\alpha_{2}=\alpha_{1}$.

The susceptibility $\chi_{S} = \chi_{0}^{zz}$ 
resulting from Eq.~(\ref{susc_eq}) 
is given by $\chi_{S}=2 C_{10}/\Delta$. The correlators $C_{\bm{R}}^{(0)zz}$ 
are calculated from Eqs.~(\ref{crz}) and (\ref{cqz}) with 
$\langle S^z \rangle^2 $ replaced by $ \displaystyle{C^{zz} \equiv \frac{1}{N} 
\sum_{\bm{R}} C_{\bm{R}}^{(0)zz}}$ (see Refs.~\onlinecite{WI97}, 
\onlinecite{ST91}), where 
the condensation part $C^{zz}$ describes long-range order (LRO). At $T=0$ we 
have the exact result $ C_{\bm{R} \neq 0}^{(0)zz} = \frac{1}{3} S^2 $. The 
ferromagnetic LRO is reflected in the divergence of $\chi_{S}$, so that 
$ \Delta(0) =\frac{4}{3} S(S+1) + \frac{4}{3} S^{2} 
\{ \lambda - (z+1) \alpha_{1} + (z-1) \alpha_{2}\} =0 $ and 
$\omega_{\bm{q}}=z \sqrt{2 \alpha_{1}/3} S (1-\gamma_{\bm{q}})$. Then, by 
Eq.~(\ref{crz}) we get 
$C_{\bm{R}}^{(0)zz}(0)=S/ \sqrt{6\alpha_{1}}\delta_{\bm{R},0} + C^{zz}$ 
resulting in the sum rule [$\bm{R}=0$, cf. Eq.~(\ref{id})]
$\frac{1}{3} S(S+1)= S/ \sqrt{6 \alpha_{1}} + C^{zz} $, and in 
$C^{zz}=\frac{1}{3} S^2$ $(\bm{R} \neq 0)$. 
Finally, at $T=0$ we obtain $\alpha_{1}(0)=3/2$ and 
$\lambda(0)+(z-1) \alpha_{2}(0)=\frac{1}{2} (3 z +1)-1/S$. 
For $S=1/2$ ($\lambda=0$) we have $\alpha_{2}(0)=\alpha_{1}(0)=3/2$.
At finite temperatures there is no LRO in the 1D and 2D systems implying 
$C^{zz}=0$. The higher sum rule (\ref{hsr}) for $n=0$ or, equivalently, 
Eq.~(\ref{u}) turns out to be trivially fulfilled. Therefore, following 
Ref.~\onlinecite{SSI94}, we put $\alpha_{2}=\alpha_{1} \equiv \alpha$ 
and $\lambda(T)=\lambda(0)=2-1/S $ and determine 
$\alpha(T)$ from the sum rule $C_{0}^{(0)zz}=\frac{1}{3}S(S+1)$. 

\section{Quantum Monte Carlo simulations}\label{mcsim}

In order to assess the accuracy of the approximations employed in the 
Green-function theory presented in the previous section we
perform QMC simulations. 
The Heisenberg ferromagnets with $S=1/2$ and $S=1$ placed on
chains or square lattices with periodic boundary conditions  
are simulated using the stochastic series expansion (SSE)
 method,\cite{sse1,sse2}
which utilizes the high-temperature series expansion
 \begin{eqnarray}
 Z = {\rm Tr}\,e^{-\beta H} = \sum_{\alpha} \sum_{n=0}^{\infty} 
 \frac{\beta^n}{n!}\left<\alpha\right| (-H)^n \left|\alpha \right>,
 \label{partfunct}
 \end{eqnarray}
where the first sum is over a complete set of states $\left| \alpha
 \right>$, 
usually taken as the eigenvectors of the $S_i^z $ operator. By
 decomposing the
Hamiltonian into diagonal and off-diagonal bond operators, introducing
 constant
unit operators to assure positivity, and reexpanding (\ref{partfunct}),
 one
finally ends up with a non-local loop representation which allows very 
efficient sampling.\cite{sse1,sse2} To minimize the effect of
 self-crossing 
and back-tracking, the directed loop-updating scheme is employed.

After initial thermalization with about $10^6$ Monte Carlo steps, 
the measurements are made after each step. During the simulation, 
the energy, magnetization, and correlation functions are 
measured and stored in a time series file, from which the specific heat 
and magnetic susceptibility can be computed using the
 fluctuation-dissipation 
relation. The correlation lengths are extracted from the exponential
falloff of the correlation functions and for comparison also by means
 of
the second-moment method.\cite{janke-greifswald06} 
Only for correlations smaller than one lattice 
spacing, small systematic deviations are visible. All those observables can 
be easily expressed by states of the 
spins on the lattice and the number and types of operators.\cite{sse3}
 The 
whole simulation usually takes of the order of $10^7$ Monte Carlo
 steps. The 
statistical error bars are estimated by the Jackknife
 method.\cite{Efron}

The results presented in this paper are generated for $S=1/2$ chains
of length up to $L=1024$ and for $S=1$ up to $L=64$. In two dimensions
 we simulate 
square lattices of edge length up to $L=64$. By comparing the 
results for different lattice sizes we made sure that for the 
 investigated 
range of temperatures and fields, the thermodynamic limit of the
 considered 
observables lies within the statistical error bars of the numerical
 results.

\section{RESULTS} \label{sec:res}
As described in Sec.~\ref{gfth}, the quantities of the Green-function theory 
determining the thermodynamic properties have to be calculated numerically 
as solutions of a coupled system of non-linear algebraic self-consistency 
equations. To this end, we use Broyden's method\cite{numrec} which yields the 
solutions with a relative error of about $10^{-7}$ on the average, where  
the numerical error increases with decreasing field and temperature. 
The momentum integrals occurring in the self-consistency equations are done by 
Gaussian integration.
Considering the $S=1/2$ ferromagnet, in Refs.~\onlinecite{JIR04} and 
\onlinecite{APP07} the thermodynamic quantities, except for the transverse 
and longitudinal correlation lengths, are calculated. Therefore, we present 
only some results for $S=1/2$ (see Figs.~\ref{fig_chi_sphalf}, 
\ref{fig_chi_max}, and \ref{fig_c_sphalf_1D}) 
which visibly improve those of Ref.~\onlinecite{JIR04}.

\subsection{Magnetic susceptibility}

% %fig_2
% \begin{figure}[b]
% \centering
% \includegraphics*[scale=0.581]{fig_2.eps} 
% \caption{Zero-field susceptibility of the 1D $S=1/2$ ferromagnet. The 
% results of the Green-function theory (solid line) and the 
% QMC data ($+:L=256$, $\times : L=1024$) are compared with the 
% QMC data of Ref.~\onlinecite{Kop89} ($\blacksquare$) and the Bethe-ansatz 
% results of Ref.~\onlinecite{YT86} ($\circ$). 
% In the inset the finite-size scaling of the 
% zero-temperature limit of $\chi_S T^2$ calculated by QMC is depicted. 
% The dashed line shows the least-square fit of the data by a linear dependence.}
% \label{fig_chi_t2}
% \end{figure}
%
%
%
%fig_2
\begin{figure}[t]
\centering
\includegraphics*[scale=0.581]{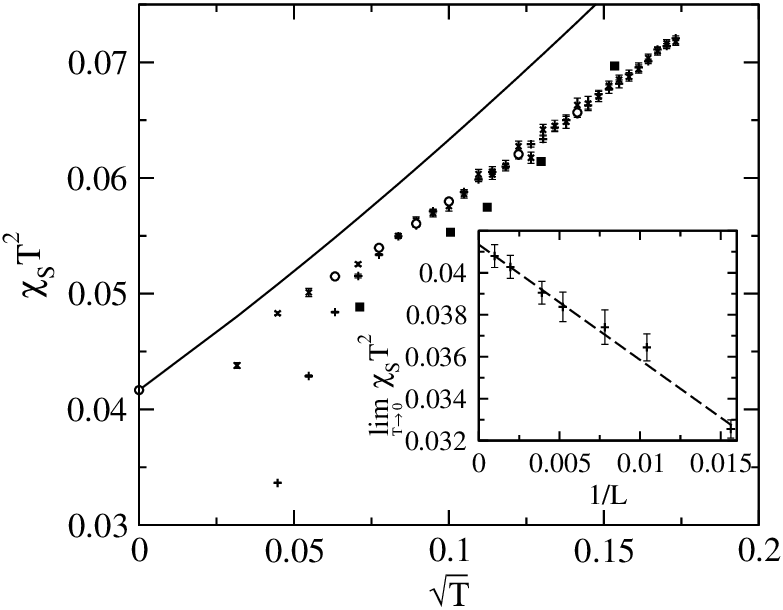} 
\caption{Zero-field susceptibility of the 1D $S=1/2$ ferromagnet. The 
results of the Green-function theory (solid line) and the 
QMC data ($+:L=256$, $\times : L=1024$) are compared with the 
QMC data of Ref.~\onlinecite{Kop89} ($\blacksquare$) and the Bethe-ansatz 
results of Ref.~\onlinecite{YT86} ($\circ$). 
In the inset the finite-size scaling of the 
zero-temperature limit of $\chi_S T^2$ calculated by QMC is depicted. 
The dashed line shows the least-square fit of the data by a linear dependence.}
\label{fig_chi_t2}
\end{figure}
%
%
% % fig_3
\begin{figure}[b]
\centering
\includegraphics*[scale=0.581]{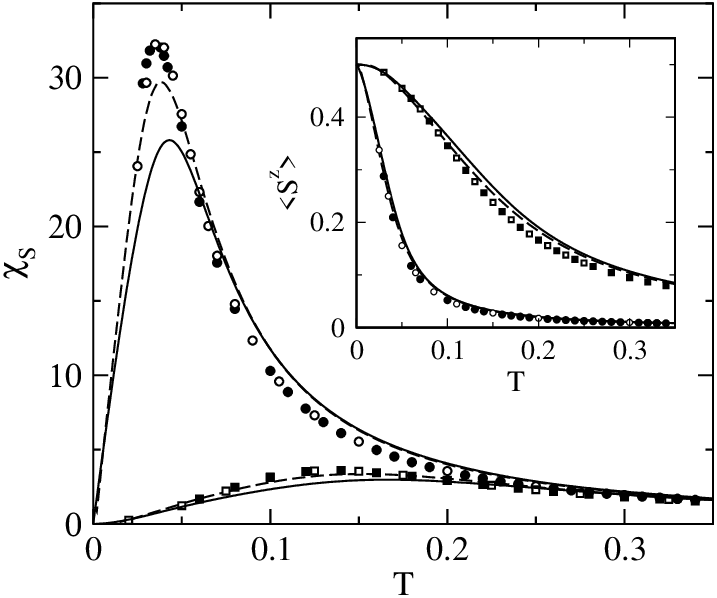}
\caption{Susceptibility of the 1D $S=1/2$ ferromagnet at
$h=0.005$ and 0.05, from top to bottom, where the results of the 
Green-function theory (solid lines) and of the Green-function method of 
Ref.~\onlinecite{APP07} (dashed lines), 
the QMC data (filled symbols, $L=128$), and the Bethe-ansatz results of
Ref.~\onlinecite{JIR04} (open symbols) are shown.
In the inset the 1D
magnetization at $h=0.005$ and 0.05, from bottom to top, is depicted.
 }
\label{fig_chi_sphalf}
\end{figure}
Let us first consider the susceptibility $\chi_S$  
in the case $h=0$, $\chi_S=2 C_{10}/\Delta$ ( see Sec.~\ref{gfth}). 
In one dimension, the low-temperature expansion yields 
$\displaystyle {\lim_{T \to 0}} \chi_{S} T^2 = \frac{2}{3} S^4 $ 
(Ref.~\onlinecite{SSI94}). Note that this result agrees with that obtained 
by the modified spin-wave theory (MSWT).\cite{Tak86} For $S=1/2$ we have 
$\displaystyle {\lim_{T \to 0}} \chi_{S} T^2 = 0.041667 $ which is in very good 
agreement with the Bethe-ansatz value 
$\displaystyle {\lim_{T \to 0}} \chi_{S} T^2 = 0.041675 $ 
(Ref.~\onlinecite{YT86}). On the other hand, previous QMC simulations by 
Handscomb's method on an $N=256$ chain combined with a renormalization-group 
approach\cite{Kop89} yield $\displaystyle {\lim_{T \to 0}} \chi_{S} T^2=0.0329$ 
(note that $\chi$ plotted in Ref.~\onlinecite{Kop89} and defined in  
Ref.~\onlinecite{KC89} is related to $\chi_S$ by $\chi = 3 \chi_S/S^2$). 
To resolve the discrepancy between the QMC results of Ref.~\onlinecite{Kop89} 
and the Bethe-ansatz value, we perform QMC simulations for chains up to 
$N=1024$ sites. The results at very low temperatures are shown in 
Fig.~\ref{fig_chi_t2} (taking the same 
plot as in Ref.~\onlinecite{Kop89}) and compared with the 
Bethe-ansatz data,\cite{YT86} the QMC data of Ref.~\onlinecite{Kop89}, and 
with the Green-function theory. Above a characteristic 
temperature, which decreases with increasing chain length, our QMC data 
agree very well with the Bethe-ansatz results. On the contrary, the QMC 
results of Ref.~\onlinecite{Kop89} for $\chi_S T^2$ are lower than ours 
by 4\% on the average. To determine the limit 
$\displaystyle {\lim_{T \to 0}} \chi_{S} T^2 $ from our QMC data, we perform a 
finite-size scaling analysis. To this end, for each chain length we linearly 
extrapolate the low-temperature linear part of the curve $ \chi_{S} T^2 $ 
to $T=0$ and fit the limiting values as function of $1/N$ by a linear 
dependence (see inset of Fig.~\ref{fig_chi_t2}). The extrapolation to $1/N=0$ 
yields $\displaystyle{\lim_{N \to \infty} \lim_{T \to 0}} \chi_S T^2 =
0.0413 \pm 0.0005$ which agrees, within the given statistical error, with 
the Bethe-ansatz value.
% %
% %fig_2
% \begin{figure}[t]
% \centering
% \includegraphics*[scale=0.581]{fig_2.eps} 
% \caption{Zero-field susceptibility of the 1D $S=1/2$ ferromagnet. The 
% results of the Green-function theory (solid line) and the 
% QMC data ($+:L=256$, $\times : L=1024$) are compared with the 
% QMC data of Ref.~\onlinecite{Kop89} ($\blacksquare$) and the Bethe-ansatz 
% results of Ref.~\onlinecite{YT86} ($\circ$). 
% In the inset the finite-size scaling of the 
% zero-temperature limit of $\chi_S T^2$ calculated by QMC is depicted. 
% The dashed line shows the least-square fit of the data by a linear dependence.}
% \label{fig_chi_t2}
% \end{figure}
% %
% %
% % % fig_3
% \begin{figure}[b]
% \centering
% \includegraphics*[scale=0.581]{fig_3.eps}
% \caption{Susceptibility of the 1D $S=1/2$ ferromagnet at
% $h=0.005$ and 0.05, from top to bottom, where the results of the 
% Green-function theory (solid lines) and of the Green-function method of 
% Ref.~\onlinecite{APP07} (dashed lines), 
% the QMC data (filled symbols, $L=128$), and the Bethe-ansatz results of
% Ref.~\onlinecite{JIR04} (open symbols) are shown.
% In the inset the 1D
% magnetization at $h=0.005$ and 0.05, from bottom to top, is depicted.
%  }
% \label{fig_chi_sphalf}
% \end{figure}
% %
% %

The 2D zero-field susceptibility in the second-order Green-function theory 
increases exponentially for $T \to 0$, $\chi_S \propto \exp(2 \pi S^2 /T) $ 
(Ref.~\onlinecite{SSI94}), where the exponent is smaller by a factor 
of two as compared with that found in the MSWT\cite{Tak86} and in the 
renormalization-group approach.\cite{KC89}

% %
% %
% % % fig_3
% \begin{figure}[b]
% \centering
% \includegraphics*[scale=0.581]{fig_3.eps}
% \caption{Susceptibility of the 1D $S=1/2$ ferromagnet at
% $h=0.005$ and 0.05, from top to bottom, where the results of the 
% Green-function theory (solid lines) and of the Green-function method of 
% Ref.~\onlinecite{APP07} (dashed lines), 
% the QMC data (filled symbols, $L=128$), and the Bethe-ansatz results of
% Ref.~\onlinecite{JIR04} (open symbols) are shown.
% In the inset the 1D
% magnetization at $h=0.005$ and 0.05, from bottom to top, is depicted.
%  }
% \label{fig_chi_sphalf}
% \end{figure}
% %
% %
%
% 
%  fig_4
\begin{figure}[t]
\centering
\includegraphics*[scale=0.581]{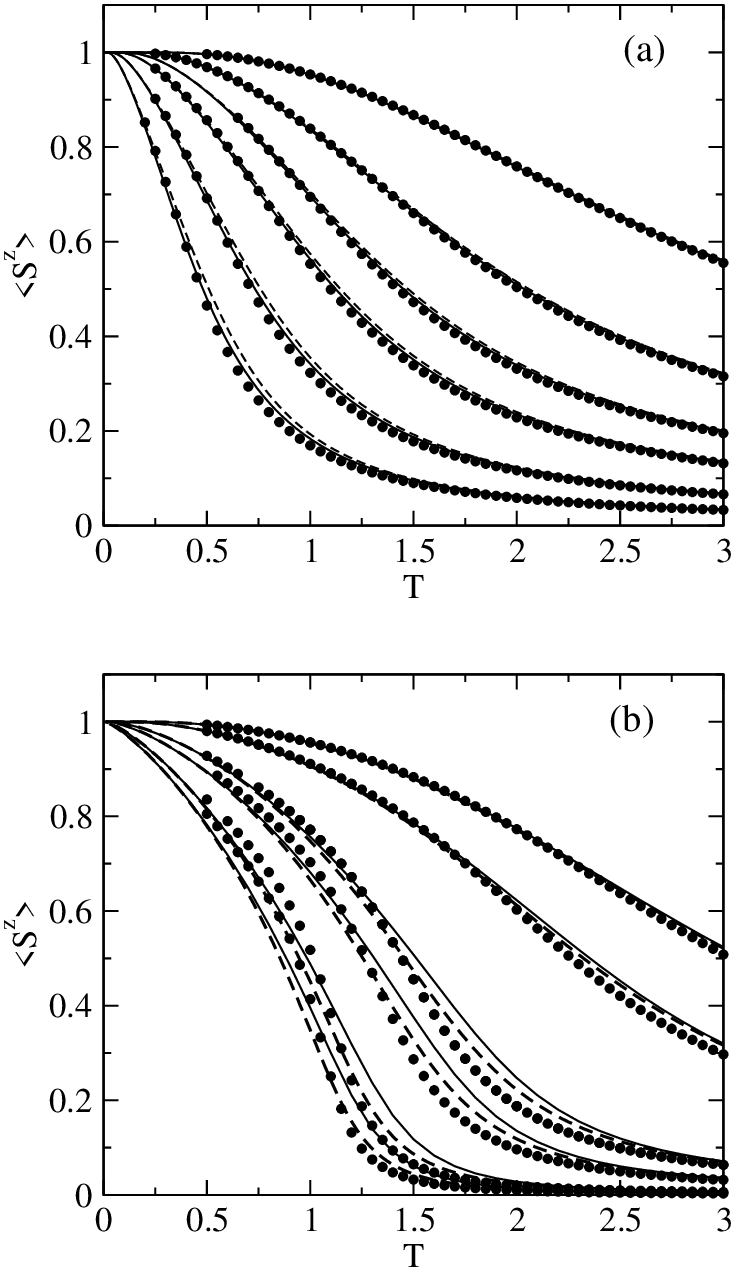}
\caption{Magnetization of the (a) 1D and (b) 2D $S=1$ ferromagnet
in magnetic fields of strengths (a) $h=0.1,0.2,0.4,0.6,1.0$, and 2.0, 
from bottom to top and (b) $h=0.005, 0.01, 0.05,0.1,0.5$, and 1.0, 
from bottom to top, as obtained by the Green-function theory 
(solid lines) and the QMC method for $L=64 $ ($\bullet$), compared with 
RPA results (dashed lines).}
\label{fig_sz_sp1}
\end{figure}
Now we consider nonzero fields and calculate the susceptibility 
$ \chi_S= \partial \langle S^z \rangle / \partial h$. First we show the 
magnetization. For $S=1/2$, as an example, $\langle S^z \rangle$ in the 1D 
model is depicted in the inset of Fig.~\ref{fig_chi_sphalf}. For the $S=1$ 
ferromagnet our analytical and QMC results in comparison with the RPA are 
plotted in Fig.~\ref{fig_sz_sp1}. 
Let us emphasize the excellent 
agreement of the theory for the chain (Fig.~\ref{fig_sz_sp1}(a)) with the QMC 
data over the whole temperature and field regions. For the 1D ferromagnet 
the RPA is a remarkably good approximation for $\langle S^z \rangle$,
as was also found in the case $S=1/2$.\cite{JIR04} In two dimensions
(Fig.~\ref{fig_sz_sp1}(b)), as compared with the QMC data, the results of our 
theory at higher temperatures are somewhat worse than those of the RPA. 
This is in contrast to the 2D $S=1/2$ ferromagnet 
for which we obtain slightly better results than the RPA at all temperatures 
and fields (improving our previous findings\cite{JIR04}).

% %
% % 
% %  fig_4
% \begin{figure}[t]
% \centering
% \includegraphics*[scale=0.581]{fig_4.eps}
% \caption{Magnetization of the (a) 1D and (b) 2D $S=1$ ferromagnet
% in magnetic fields of strengths (a) $h=0.1,0.2,0.4,0.6,1.0$, and 2.0, 
% from bottom to top and (b) $h=0.005, 0.01, 0.05,0.1,0.5$, and 1.0, 
% from bottom to top, as obtained by the Green-function theory 
% (solid lines) and the QMC method for $L=64 $ ($\bullet$), compared with 
% RPA results (dashed lines).}
% \label{fig_sz_sp1}
% \end{figure}
% % 
% %
% % fig_5
\begin{figure}[t]
\centering
\includegraphics*[scale=0.581]{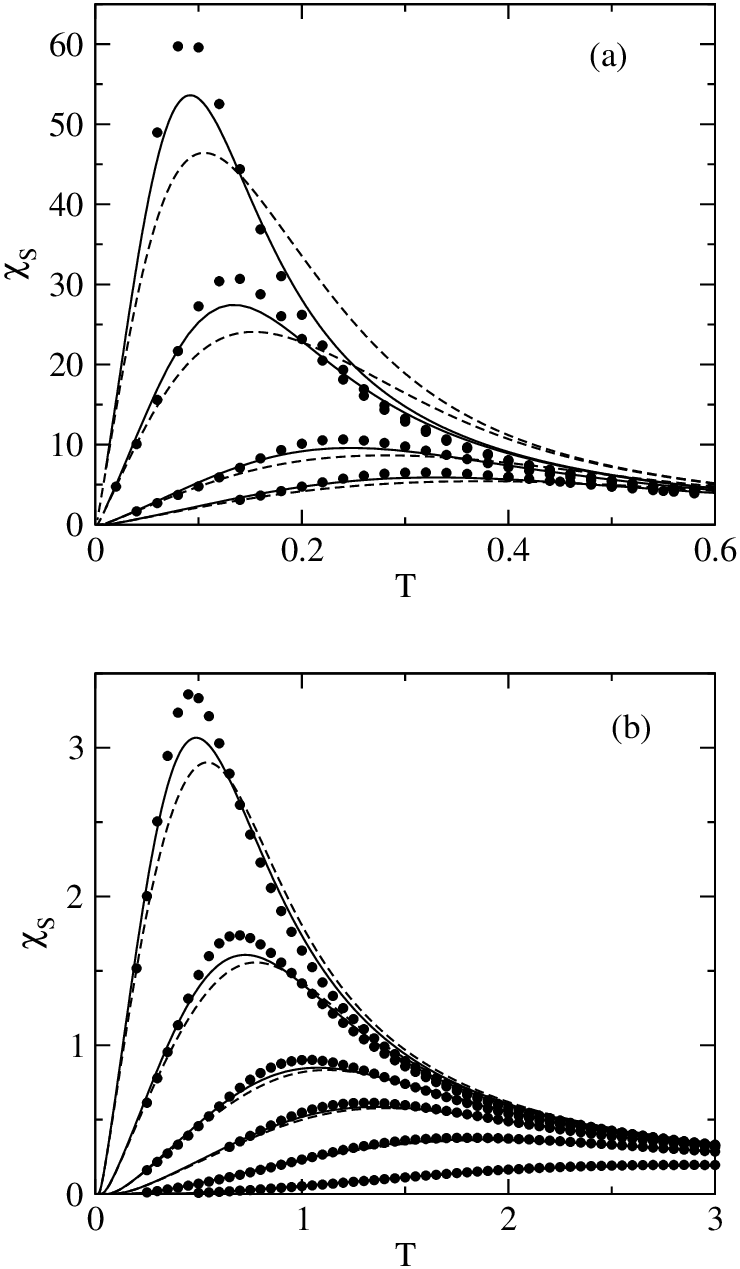}
\caption{Susceptibility of the 1D $S=1$ ferromagnet (a) at low fields,  
$h=0.005, 0.01,0.03$, and 0.05, from top to bottom, and (b) at higher fields,  
$h=0.1, 0.2, 0.4, 0.6, 1.0$, and 2.0, from top to bottom, 
where the Green-function (solid lines), the QMC ($\bullet$, $L=64$),  
and the RPA results (dashed lines) are shown.}
\label{fig_chi_sp1_1D}
\end{figure}
\begin{figure}[t]
\centering
\includegraphics*[scale=0.581]{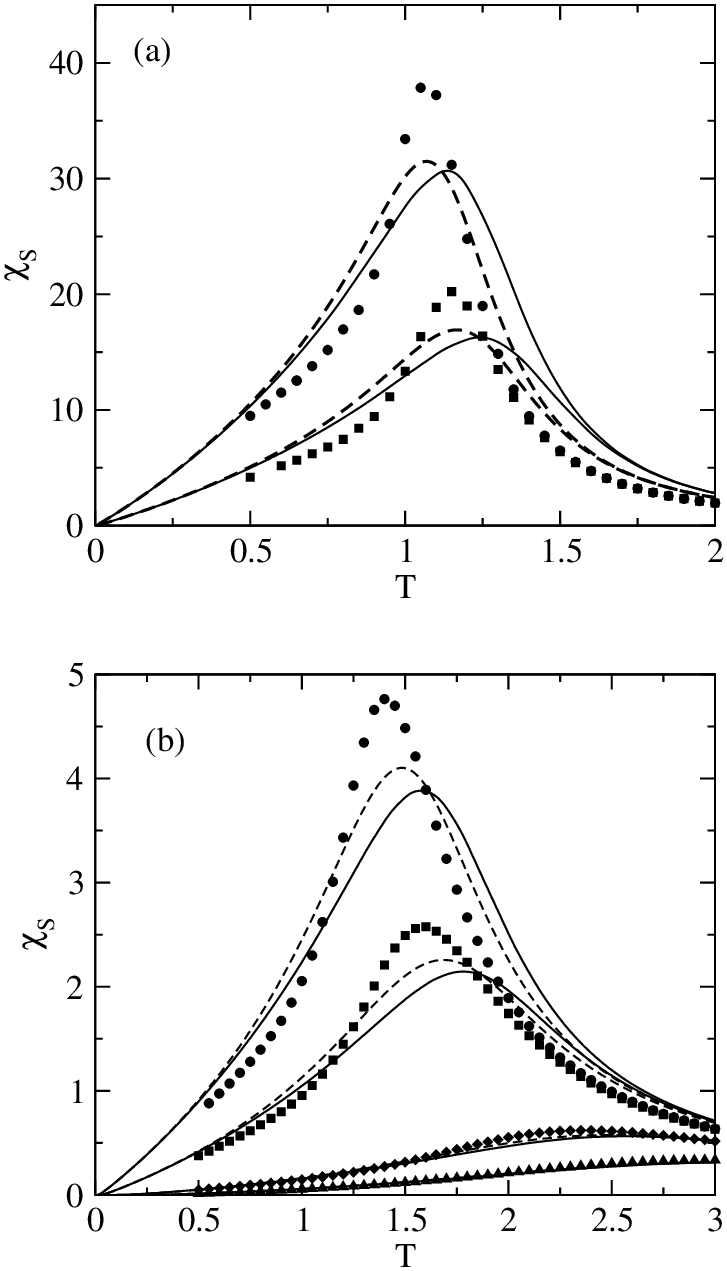}
\caption{Susceptibility of the 2D $S=1$ ferromagnet (a) at very
low fields, $h=0.005$ and 0.01, from top to bottom, and (b) at higher 
fields, $h=0.05, 0.1, 0.5$, and 1.0, from top to bottom, 
obtained by the Green-function theory (solid lines),
QMC for $L=64$ (filled symbols), and RPA (dashed lines).}
\label{fig_chi_sp1_2D}
\end{figure}

The susceptibility for $h \neq 0$ vanishes at $T=0$. 
Therefore, $\chi_S (T)$ has a maximum at 
$T_m^{\chi}$, where $T_m^{\chi}$ increases and the height of the 
susceptibility maximum $\chi_S (T_m^{\chi})$ decreases with increasing field. 
For $S=1/2$, in Fig.~\ref{fig_chi_sphalf} the low-field susceptibility
in the 1D model 
is shown, where for $h=0.005$ a better agreement of the theory with the 
Bethe-ansatz results is found than in Ref.~\onlinecite{JIR04}. Note that 
our QMC data are in a very good agreement with the Bethe results. For 
comparison, in Fig.~\ref{fig_chi_sphalf} the susceptibility in the simplified 
approach 
with $\alpha_{2}^{\nu \mu}=\alpha_{1}^{\nu \mu}$ (Ref.~\onlinecite{APP07}), 
where the equality (\ref{susc_eq}) is disregarded and the regularity condition 
(\ref{rb1}) is used instead of the higher sum rule (\ref{hsr_05}), is plotted 
as well. It is remarkable that $\chi_S$ in this approach is in a better 
agreement with the exact methods than the susceptibility in our extended 
theory with $\alpha_{2}^{\nu \mu} \neq \alpha_{1}^{\nu \mu}$. However, 
considering the correlation length the situation changes qualitatively (see 
below).
For $S=1$ the 
susceptibility is plotted in Figs.~\ref{fig_chi_sp1_1D} and 
\ref{fig_chi_sp1_2D}. In one dimension (Fig.~\ref{fig_chi_sp1_1D}),
the good agreement between Green-function theory and QMC
corresponds to the results depicted in Fig.~\ref{fig_sz_sp1}(a). As compared
with the QMC data for the 2D model (Fig.~\ref{fig_chi_sp1_2D}), in RPA the 
maximum position $T_{m}^{\chi}$ is somewhat better reproduced than in our 
theory. 

% 
%  fig_6
% 
% 
%  fig_7
\begin{figure}[h]
\centering
\includegraphics*[scale=0.581]{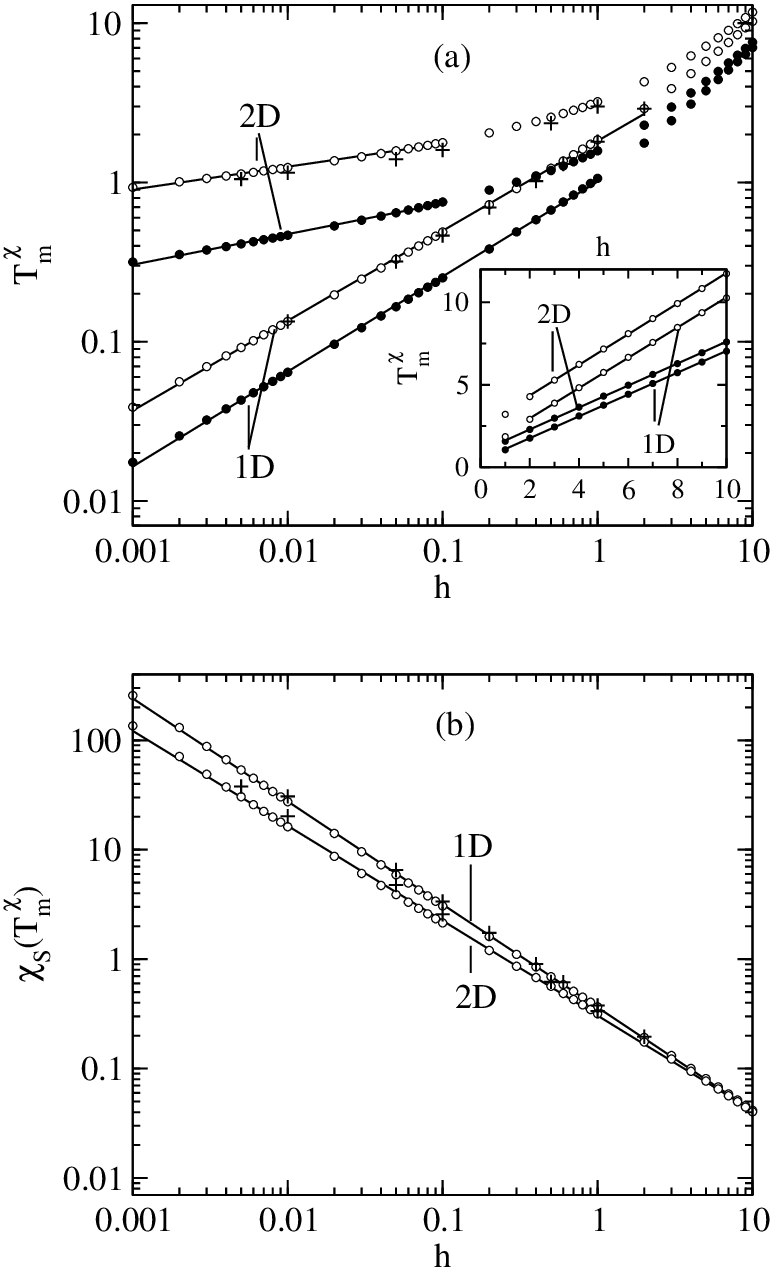}
\caption{Field dependence of the (a) position and (b) height of the 
susceptibility maximum obtained by the Green-function theory 
for the $S=1/2$ ($\bullet$) and $S=1$ ($\circ$) ferromagnets and fit 
by power laws (solid lines) in comparison with the QMC data (+, $S=1$, $L=64$). 
The inset shows  the fit of $T_m^{\chi}$ at high 
fields by a linear dependence. For clarity, $\chi_S(T_m^{\chi})$ is plotted 
for $S=1$ only.}
\label{fig_chi_max}
\end{figure}

To analyze the field dependence of $T_{m}^{\chi}$ and 
$\chi_{S}(T_{m}^{\chi})$ in more detail as in our previous 
paper,\cite{JIR04} the calculations are extended to a much
broader field region, $0.001 \leqslant h \leqslant 10$. As can be seen in
Fig.~\ref{fig_chi_max}(a), at low fields the theory may be well fitted by the 
power law
\begin{equation}
T_{m}^{\chi}=a h^{\gamma},
\label{pl}
\end{equation}
where the field regions and the values of $a$ and $\gamma$ are given in
Table \ref{tab_1}. Let us point out that the theory for the 1D 
$S=1/2$ model is in reasonable 
agreement with the Bethe-ansatz result at $h \leqslant 0.1$,\cite{JIR04}
$a=0.765$ and $\gamma=0.576$. In the high-field region, 
$T_{m}^{\chi}$ obeys a linear dependence 
(cf.~inset of Fig.~\ref{fig_chi_max}(a)),
\begin{equation}
T_{m}^{\chi}=\tilde{a} h + \tilde{b}
\label{ll}
\end{equation}
with $\tilde{a}$ and $\tilde{b}$ given in Table \ref{tab_1}. Note that the 
linear law (\ref{ll}) was not found in Ref.~\onlinecite{JIR04}. 
Our results for the maximum height
$\chi_{S}(T_{m}^{\chi})$ as a function of $h$ may be well described in the
whole field region $0.001 \leqslant h \leqslant 10.0$ 
(see Fig.~\ref{fig_chi_max}(b)) by the power law
\begin{equation}
\chi_{S} (T_{m}^{\chi})=b h^{\beta},
\label{hipl}
\end{equation}
where the coefficients  are given in Table \ref{tab_2}. The values of 
$b$ and $\beta$ for $S=1/2$ slightly deviate (by about 5\% on the average) from 
those found previously.\cite{JIR04} Again, our theory for $S=1/2$ 
is in reasonable agreement with the 1D Bethe-ansatz result at 
$h \leqslant 0.1$, $b=0.208$ and $\beta=-0.952$ (Ref.~\onlinecite{JIR04}).

For comparison, we consider the power-law behavior in RPA. We find the RPA 
results in the low- and high-field regions to be well fitted by the laws 
(\ref{pl})-(\ref{hipl}), where the coefficients are in good agreement with the 
values given in Tables \ref{tab_1} and \ref{tab_2}. More precisely, for 
the 1D and 2D $S=1/2$ and $S=1$ models the average deviations of the 
coefficients in the laws (\ref{pl}), (\ref{ll}), and (\ref{hipl}) amount to 
about 6\%, 3\%, and 2\%, respectively. For example, considering the $S=1/2$ 
ferromagnet in high fields, $2 \leqslant h \leqslant 10$, we obtain the linear 
dependence (\ref{ll}) for the 1D (2D) case with $\tilde{a}=0.657$ (0.661) 
and $\tilde{b}=0.496$ (1.015) which yields a better fit than the power law 
(\ref{pl}). Recently, in Ref.~\onlinecite{HCP07} such a law was given for 
the 1D (2D) model in the region $3 \; (4.4) \leqslant h \leqslant 6.5$. 
Even in this limited field region, we find the fit by the linear law 
(\ref{ll}) to be slightly better than the fit by the power law (\ref{pl}) 
(see Ref.~\onlinecite{HCP07}).

% \begin{widetext}
% % table 1
% \begin{table}[t]
% \caption{Validity regions ($h$) and coefficients of the power laws 
% (\ref{pl}) and (\ref{ll}) 
% for the susceptibility of the
% 1D and 2D $S=1/2$ and $S=1$ ferromagnets.}
% \begin{center}
% \begin{tabular}{ccc|cc} \hline \hline
% & \multicolumn{2}{c|}{$S=1/2$} & \multicolumn{2}{c}{$S=1$}  \\ \cline{2-5}
% \phantom{mmm}& 1D        & 2D        & 1D        & 2D        \\ \hline
% $h$       & \phantom{m} $0.001 - 1.0$ \phantom{m}
% & \phantom{m} $0.001 - 0.1$ \phantom{m} & \phantom{m} $0.001 - 2.0$
% \phantom{m} & \phantom{m} $0.001 - 0.1$ \phantom{m} \\
% $a$       & 1.013 & 1.149 & 1.823 & 2.433   \\
% $\gamma $ & 0.596 & 0.192 & 0.565 & 0.144   \\ \hline
% $h$       & $1.0 - 10.0$  & $1.0 - 10.0$  & $2.0 - 10.0$  & $2.0 - 10.0$  \\
% $\tilde{a}$ &  0.661 & 0.666 & 0.917 & 0.929  \\
% $\tilde{b}$ &  0.443 & 0.961 & 1.136 & 2.494  \\ \hline \hline
% \end{tabular}
% \end{center}
% \label{tab_1}
% \end{table}
% \end{widetext}
% %
% table 1
\begin{table}[t]
\caption{Validity regions ($h$) and coefficients of the power laws 
(\ref{pl}) and (\ref{ll}) 
for the susceptibility of the
1D and 2D $S=1/2$ and $S=1$ ferromagnets.}
\begin{center}
\begin{tabular}{ccc|cc} \hline \hline
& \multicolumn{2}{c|}{$S=1/2$} & \multicolumn{2}{c}{$S=1$}  \\ \cline{2-5}
\phantom{mmm}& 1D        & 2D        & 1D        & 2D        \\ \hline
$h$       & $0.001 - 1.0$ 
& \phantom{l} $0.001 - 0.1$ \phantom{l} &  \phantom{l} $0.001 - 2.0$ \phantom{l}
 & $0.001 - 0.1$  \\
$a$       & 1.013 & 1.149 & 1.823 & 2.433   \\
$\gamma $ & 0.596 & 0.192 & 0.565 & 0.144   \\ \hline
$h$       & $1.0 - 10.0$  & $1.0 - 10.0$  & $2.0 - 10.0$  & $2.0 - 10.0$  \\
$\tilde{a}$ &  0.661 & 0.666 & 0.917 & 0.929  \\
$\tilde{b}$ &  0.443 & 0.961 & 1.136 & 2.494  \\ \hline \hline
\end{tabular}
\end{center}
\label{tab_1}
\end{table}
%
%
% table 2
\begin{table}[t] % [htp]-vel a kovetkezo cim fele megy
\caption{Coefficients of the power law (\ref{hipl}) 
for the susceptibility of the 1D and 2D $S=1/2$ 
and $S=1$ ferromagnets in the field region $0.001 \leqslant h \leqslant 10.0$}.
\begin{center}
\begin{tabular}{ccc|cc} \hline \hline
& \multicolumn{2}{c|}{$S=1/2$} & \multicolumn{2}{c}{$S=1$}   \\ \cline{2-5}
\phantom{mmm}&  1D    &  2D       &  1D       &  2D        \\ \hline
$b$       &  \phantom{m} 0.192 \phantom{m}    &  \phantom{m} 0.166 \phantom{m}
   & \phantom{m} 0.362 \phantom{m}  & \phantom{m} 0.305 \phantom{m}     \\
$\beta $  & $-0.925$ & $-0.850$ & $-0.941$ & $-0.867$     
\\ \hline \hline
\end{tabular}
\end{center}
\label{tab_2}
\end{table}

\subsection{Correlation length}

To obtain the transverse and longitudinal correlation lengths $\xi^{+-}$ and 
$\xi^{zz}$ we consider the long-distance correlators  
$C_{\bm{R}}^{(0)-+}$ and 
$\bar{C}_{\bm{R}}^{(0)zz} \equiv C_{\bm{R}}^{(0)zz} - 
\langle S^z \rangle^2$ with $C_{\bm{R}}^{(0)zz}$ calculated by 
Eq.~(\ref{crz}), respectively. Note that the temperature dependence of both 
$C_{\bm{R}}^{(0)-+}$ and $\bar{C}_{\bm{R}}^{(0)zz}$ exhibits a maximum, because 
the correlators vanish at $T=0$, following from Eqs.~(\ref{crz}) and 
(\ref{crpm}), and for $T \to \infty$. By the asymptotic ansatz 
\begin{equation}
C_{\bm{R}}^{(0)-+}=A^{+-} \exp ( -R/\xi^{+-}),
\label{crp_0}
\end{equation} 
\begin{equation}
\bar{C}_{\bm{R}}^{(0)zz}=A^{zz}\exp( -R/\xi^{zz}), 
\label{crz_0}
\end{equation} 
and the logarithmic plot of the correlators as functions of 
$R=|\bm{R}|$ the inverse correlation lengths are evaluated numerically 
from linear fits.

In the literature, often the correlation length is determined from the 
expansion of the static spin susceptibility around the magnetic wavevector 
(see, e.g., Refs.~\onlinecite{ST91}, \onlinecite{WI97}, 
and \onlinecite{JIR05}). In the 
ferromagnetic case we expand the static susceptibilities 
$\chi_{\bm{q}}^{+-}$ (resulting from Eqs.~(\ref{omqp2})-(\ref{gf}), 
(\ref{omq12}), and (\ref{aq12})) and $\chi_{\bm{q}}^{zz}$ (given by 
Eqs.~(\ref{gfz})-(\ref{omqz})) around $\bm{q}=0$, 
$\chi_{\bm{q}}^{\nu \mu} = \chi_{0}^{\nu \mu}/[1+(\xi_{\chi}^{\nu \mu})^2 q^2] 
\; (\nu \mu = +-, \; zz)$. We obtain
\begin{equation}
\xi_{\chi}^{+-}=\sqrt{\alpha_{1}^{+-} \langle S^z \rangle/h}
\label{ksip_chi}
\end{equation}
and 
\begin{equation}
\xi_{\chi}^{zz}=\sqrt{2 \alpha_{1}^{zz} C_{10}^{(0)-+}/\Delta^{zz}}.
\label{ksiz_chi}
\end{equation}
Deriving Eq.~(\ref{ksip_chi}) the regularity condition (\ref{rb1}) for 
$n=0$, which reads as 
$h \langle S^z \rangle =z C_{10} g_0$, and Eq.~(\ref{g0}), yielding the 
relation 
$\Delta^{+-}=2 h (C_{10}/\langle S^z \rangle - 
\alpha_{1}^{+-} \langle S^z \rangle)$, have been used. Let us point out that 
the correlation lengths $\xi_{\chi}^{\nu \mu}$ generally deviate from 
$\xi^{\nu \mu}$ defined by Eqs.~(\ref{crp_0}) and (\ref{crz_0}).
%
% %  fig_8
% \begin{figure}[b]
% \centering
% \includegraphics*[scale=0.581]{fig_8.eps}
% \caption{Zero-field correlation length of the 1D ferromagnet with $S=1/2$ 
% obtained by the Green-function theory (solid lines) and by QMC simulations 
% ($\times$, $L=32$) and with $S=1$ resulting from the theory (long-dashed line). 
% For comparison, the correlation length $\xi_{\chi}$ determined from 
% the expansion of 
% the static susceptibility around $\bm{q}=0$ is plotted for $S=1/2$ (dotdashed 
% line) and $S=1$ (dotted line). The results for $S=1/2$ are compared 
% with the Bethe-ansatz 
% data ($\circ$) of Ref.~\onlinecite{Yam90} and with the QMC data of 
% Ref.~\onlinecite{Kop89} ($\bullet$) depicted in the inset together with a 
% one-parameter fit (short-dashed line).}
% \label{fig_8}
% \end{figure}
% % 
% 

\begin{figure}[t]
\centering
\includegraphics*[scale=0.581]{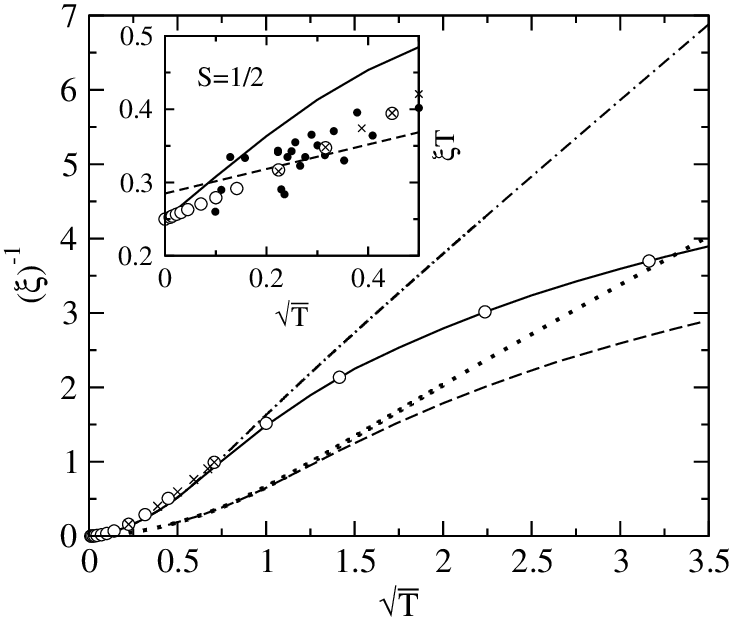}
\caption{Zero-field correlation length of the 1D ferromagnet with $S=1/2$ 
obtained by the Green-function theory (solid lines) and by QMC simulations 
($\times$, $L=32$) and with $S=1$ resulting from the theory (long-dashed line). 
For comparison, the correlation length $\xi_{\chi}$ determined from 
the expansion of 
the static susceptibility around $\bm{q}=0$ is plotted for $S=1/2$ (dotdashed 
line) and $S=1$ (dotted line). The results for $S=1/2$ are compared 
with the Bethe-ansatz 
data ($\circ$) of Ref.~\onlinecite{Yam90} and with the QMC data of 
Ref.~\onlinecite{Kop89} ($\bullet$) depicted in the inset together with a 
one-parameter fit (short-dashed line).}
\label{fig_8}
\end{figure}

First we consider the correlation length in zero field, where 
$\xi^{+-}=\xi^{zz} \equiv \xi $. In one dimension, the low-temperature 
expansion yields $\lim_{T \to 0} \xi T = S^2$ (Ref.~\onlinecite{SSI94}) 
which agrees with the MSWT result\cite{Tak86} and, for $S=1/2$, with the 
result obtained by the thermal Bethe-ansatz method of Ref.~\onlinecite{Yam90}. 
The renormalization-group approach of Ref.~\onlinecite{Kop89} combined with 
QMC simulations yields $\lim_{T \to 0} \xi T =1.14 S^2$. In Fig.~\ref{fig_8} 
the zero-field correlation length of the 1D ferromagnet is shown. Let 
us stress the very good agreement of our QMC data for $S=1/2$ with the 
Bethe-ansatz results of Ref.~\onlinecite{Yam90}. Even on the finer scale of the 
inset, deviations are almost invisible. For comparison, also the QMC data of 
Ref.~\onlinecite{Kop89} and a one-parameter fit are given in the inset. 
Moreover, we obtain a good agreement of the Green-function theory, where 
$\xi$ is calculated from the definition (\ref{crp_0}), with our QMC data. 
In addition to $\xi$, in Fig.~\ref{fig_8} the correlation length $\xi_{\chi}$ 
calculated for $S=1/2$ and $S=1$ by Eq.~(\ref{ksiz_chi}) 
[$\alpha_{1}^{zz}=\alpha$, $C_{10}^{(0)-+}=C_{10}$, $\Delta^{zz}=\Delta$ given 
by Eq.~(\ref{delta_05})] is plotted. For $T \lesssim 0.25 $, i.e. $\xi>1$, 
$\xi_{\chi}$ nearly coincides with $\xi$. With increasing 
temperature, i.e., with decreasing $\xi<1$, the deviation of $\xi_{\chi}$ 
from $\xi$ appreciably increases. In the high-temperature limit we get 
$\xi_{\chi}^{-1}=\{ 3T/S(S+1)\}^{1/2}$ resulting from  $C_{10}=2[S(S+1)]^2 /9T$ 
(Ref.~\onlinecite{SSI94}). In the following we plot $\xi_{\chi}$ in such 
cases only, where $\xi_{\chi}$ remarkably deviates from $\xi$.
%
%  fig_8

In two dimensions, the zero-field correlation length in the second-order 
Green-function theory increases exponentially for $T \to 0$, 
$\xi \propto \exp (\pi S^2/T)$ (Ref.~\onlinecite{SSI94}). As is the case 
for the magnetic susceptibility, the exponent is smaller by a factor of two 
as compared with the MSWT\cite{Tak86} and the renormalization-group 
approach.\cite{KC89}  
%
%  fig_9
\begin{figure}[t]
\centering
\includegraphics*[scale=0.581]{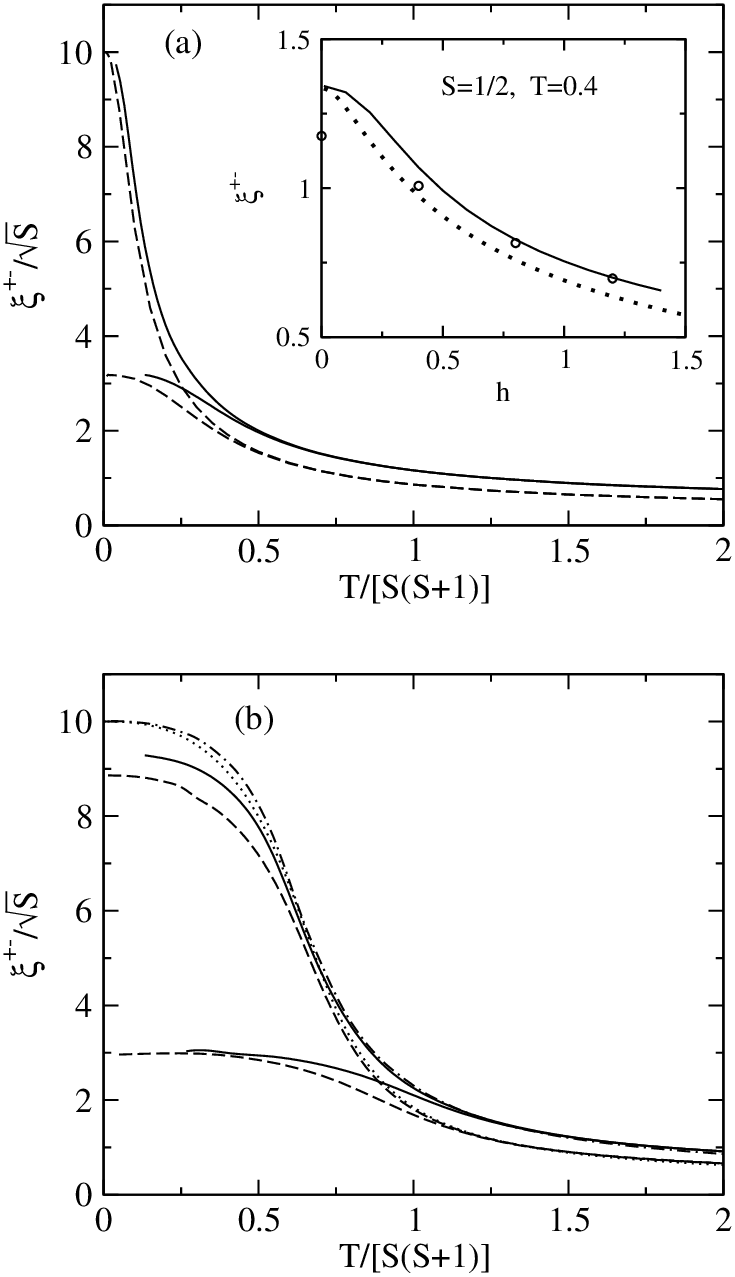}
\caption{Transverse correlation length of the (a) 1D and (b) 2D ferromagnet 
with $S=1/2$ 
(solid lines) and $S=1$ (dashed lines) in the fields $h=0.01$ and 0.1, 
from top to bottom. 
In the 2D case at $h=0.01$, the correlation length $\xi_{\chi}^{+-}$ calculated 
from the static susceptibility is shown for $S=1/2$ (dotdashed line) and $S=1$ 
(dotted line). In the inset the 
results of the Green-function theory are compared with the Bethe-ansatz data 
of Ref.~\onlinecite{Tak91} ($\circ$) and the RPA (dotted line).  
}
\label{fig_9}
\end{figure}
% 
% 
%
%
%  fig_10
\begin{figure}[t]
\centering
\includegraphics*[scale=0.581]{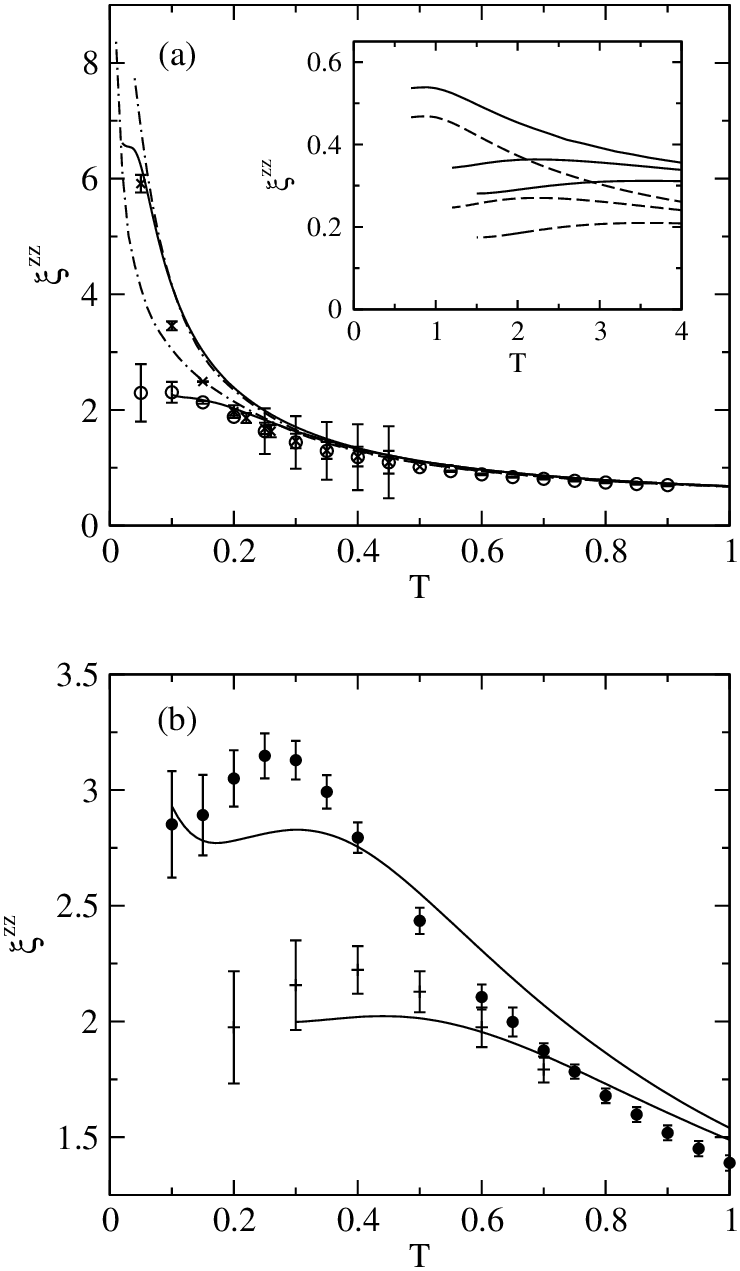}
\caption{Longitudinal correlation length of the 1D ferromagnet with (a) $S=1/2$   
and (b) $S=1$ in the fields (a) $h=0.005$ and 0.05 and (b) $h=0.05$ and 0.1, 
from top to bottom, calculated by the Green-function (solid lines) 
and QMC methods 
($\times$, $\circ$; $L=32$ and $\bullet$, +; $L=32$) and, for $S=1/2$, by the 
method of Ref.~\onlinecite{APP07} (dotdashed lines). 
The inset exhibits the results 
for $S=1/2$ at the strong fields $h=1$, 3, and 5, from top to bottom, 
in comparison 
with the correlation length $\xi_{\chi}^{zz}$ (dashed lines) obtained from the 
static susceptibility.   
}
\label{fig_10}
\end{figure}

For $h \neq 0$ the transverse and longitudinal correlation lengths reveal 
qualitatively different temperature dependences. Considering the transverse 
correlation length $\xi^{+-}$ shown in Fig.~\ref{fig_9}, the magnetic field 
cuts off the divergence of the zero-field correlation length at $T=0$ 
which corresponds to the absence of a phase transition and is evident 
from Eq.~(\ref{ksip_chi}), $\xi_{\chi}^{+-}(T=0)=\sqrt{S/h}$ agreeing with 
the RPA result (\ref{rpa_ksi}) derived in the Appendix. As can be seen in 
the inset of 
Fig.~\ref{fig_9}(a), in the 1D $S=1/2$ model we obtain a good agreement of our 
analytical results for $T=0.4$ and $h \leqslant 1.2$ with the Bethe-ansatz 
data of Ref.~\onlinecite{Tak91}. However, the comparison of the theory with the 
available Bethe data for $T=0.4$ and fields up to $h=4$ and for 
$T=0.2$ (Ref.~\onlinecite{Tak91}) is hampered by numerical uncertainties 
resulting from 
too small values of $\Delta^{+-}$. Note the remarkably good agreement of 
$\xi^{+-}$ with the RPA results (see inset). Concerning the dimensional 
dependence, in contrast to the case $h=0$, $\xi^{+-}$ in one and two dimensions 
exhibits qualitatively the same behavior as $T \to 0$. In the 2D model 
[Fig.~\ref{fig_9}(b)], the deviation of $\xi_{\chi}^{+-}$ from $\xi^{+-}$ 
increases with decreasing temperature, i.e., with increasing $\xi^{+-} > 1$ 
which is clearly seen at $h=0.01$ and is opposite to the behavior in the $h=0$ 
case.

\begin{figure}[b]
\centering
\includegraphics*[scale=0.581]{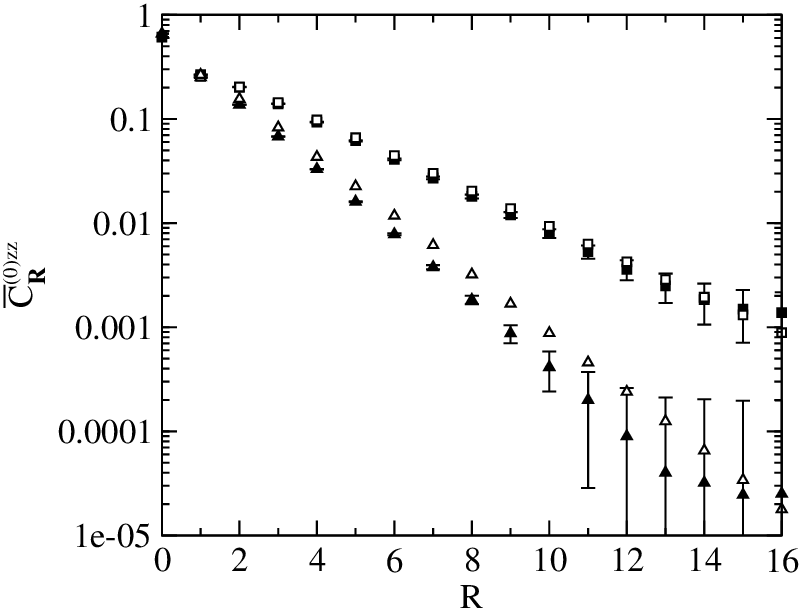}
\caption{Correlation function $\bar{C}_{\bm{R}}^{(0)zz}=
\langle S_{0}^{z} S_{\bm{R}}^{z}\rangle - \langle S^{z} \rangle^2$ vs 
$R=|\bm{R}|$ for the 1D $S=1$ ferromagnet in the field $h=0.05$ at $T=0.5$ 
and 1.0, from top to bottom, calculated by the Green-function theory 
(open symbols) and by QMC (filled symbols, $L=32$).}
\label{fig_11}
\end{figure}

In Fig.~\ref{fig_10} the longitudinal correlation length of the 1D ferromagnet 
is shown, where the QMC data are found to be in a fair agreement with 
our theory. This refers, in particular, to the $S=1/2$ model, where our  
results obtained by  the simplified approach of Ref.~\onlinecite{APP07} 
are plotted as well. Considering $h=0.05$, at low temperatures those results 
remarkably deviate from the QMC data and our extended theory with 
$\alpha_{2}^{\nu \mu} \neq \alpha_{1}^{\nu \mu}$. In contrast to $\xi^{+-}$, 
the behavior of $\xi^{zz}$ as $T \to 0$ is not conclusive which is due to 
numerical uncertainties at low temperatures, where the long-distance 
correlators $\bar{C}_{\bm{R}}^{(0)zz}$ needed to calculate $\xi^{zz}$ are very 
small. For example, for $S=1/2$ and strong fields [see inset of 
Fig.~\ref{fig_10}(a)] the relevant correlators in the temperature region, 
where results are not given, are smaller than about $10^{-10}$ to $10^{-14}$.
Moreover, for $S=1$ the results of the theory are reliable only at 
$T>T_0 \simeq 0.1$ and 0.3 for $h=0.05$ and 0.1, respectively 
[see Fig.~\ref{fig_10}(b)]. At $T < T_0$, the relevant correlators, being 
smaller than about $10^{-4}$, reveal an unreasonable behavior. This may be 
ascribed to our choice of a closed system of self-consistency equations 
for $S>1/2$, as described in Sec.~\ref{gfth}. Whereas the relative deviation 
of the NN correlators $C_{10}^{(0)zz}$ resulting from the self-consistency 
equations and from Eq.~(\ref{crz}) is small (see Sec.~\ref{gfth}), the 
corresponding deviation of the correlators $\bar{C}_{10}^{(0)zz}$ becomes 
very large at low temperatures.
Depending on the field and 
spin, the temperature dependence of $\xi^{zz}$ in the 1D ferromagnet reveals 
a maximum at $T_m^{\xi}>0$. 
This anomaly can be clearly seen in the 1D $S=1$ model at low 
fields [Fig.~\ref{fig_10}(b)]. On the other hand, in the 1D $S=1/2$ model the 
maximum appears at high fields, $h>0.8$ [see inset of Fig.~\ref{fig_10}(a)]. 
Moreover, as can be seen from Fig.~\ref{fig_10}, keeping the field $h=0.05$ 
fixed, the maximum develops with increasing spin. 
Note that a maximum of $\xi^{zz}$ at a finite temperature is not obtained by 
the approach of Ref.~\onlinecite{APP07}. To our knowledge, such an  
anomaly in the correlation length has not been found before. 
To get some insight into the maximum of $\xi^{zz}$, we first suggest that 
larger correlation lengths may be connected with larger correlation functions. 
Correspondingly, we consider the maximum of $\bar{C}_{\bm{R}}^{(0)zz}$ at 
$T_m^{zz}(R)$, where $T_m^{zz}(R) > T_m^{\xi}$. By a detailed analysis we 
find $T_m^{zz}(R)$ in the limit $R \to \infty$ to coincide with $T_m^{\xi}$ 
in all cases, where $\xi^{zz}$ has a maximum at $T_m^{\xi} > 0$ 
(see Fig.~\ref{fig_10}), i.e., 
% %
% %
% %  fig_10
% \begin{figure}[t]
% \centering
% \includegraphics*[scale=0.581]{fig_10.eps}
% \caption{Longitudinal correlation length of the 1D ferromagnet with (a) $S=1/2$   
% and (b) $S=1$ in the fields (a) $h=0.005$ and 0.05 and (b) $h=0.05$ and 0.1, 
% from top to bottom, calculated by the Green-function (solid lines) 
% and QMC methods 
% ($\times$, $\circ$; $L=32$ and $\bullet$, +; $L=32$) and, for $S=1/2$, by the 
% method of Ref.~\onlinecite{APP07} (dotdashed lines). 
% The inset exhibits the results 
% for $S=1/2$ at the strong fields $h=1$, 3, and 5, from top to bottom, 
% in comparison 
% with the correlation length $\xi_{\chi}^{zz}$ (dashed lines) obtained from the 
% static susceptibility.   
% }
% \label{fig_10}
% \end{figure}
%
$\lim_{R \to \infty} T_m^{zz}(R) = T_m^{\xi}$. This result is corroborated by 
the conditions for a maximum which may be derived from the ansatz (\ref{crz_0}). 
We get $\frac{1}{R} \partial \ln \bar{C}_{\bm{R}}^{(0)zz} / \partial T = 
\frac{1}{R} \partial \ln A^{zz} / \partial T + 
\frac{1}{\xi} \partial \ln \xi / \partial T$. At $T_m^{zz}(R)$ we have 
$\frac{1}{\xi} \partial \ln \xi / \partial T = 
- \frac{1}{R} \partial \ln A^{zz} / \partial T$ and, for $ R \to \infty$, 
$\partial \xi / \partial T =0$. As can be easily verified, the maximum 
condition $\partial^2 \bar{C}_{\bm{R}}^{(0)zz} / \partial T^2<0$ results in 
$\partial^2 \xi^{zz} / \partial T^2 <0$. To compare the 
QMC and Green-function methods yielding the anomaly of $\xi^{zz}$ in the 1D 
$S=1$ model [Fig.~\ref{fig_10}(b)] in more detail, in Fig.~\ref{fig_11} 
the distance dependence of the corresponding correlator 
$\bar{C}_{\bm{R}}^{(0)zz}$ at $h=0.05$ is depicted. For $T=0.5$ a very good 
agreement of both methods is found. 
% 
% 
%  fig_11
% 
%
% 
%  fig_12
\begin{figure}[t]
\centering
\includegraphics*[scale=0.581]{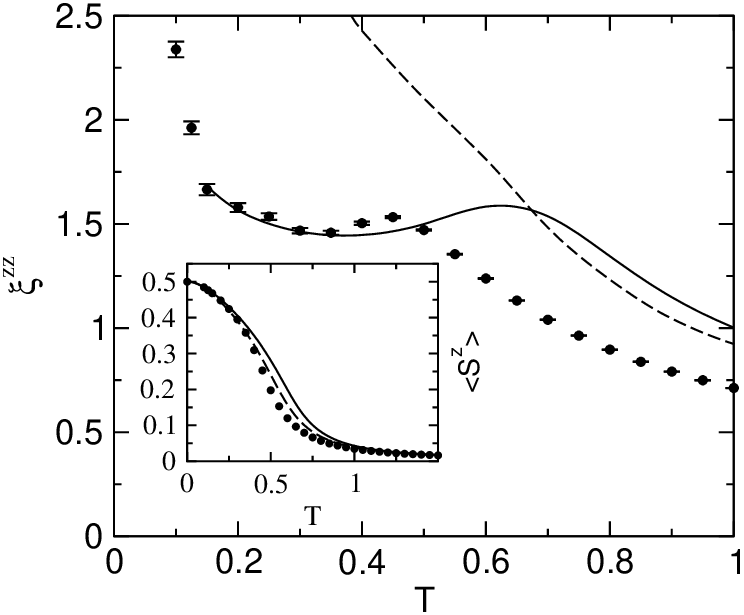}
\caption{Longitudinal correlation length of the 2D $S=1/2$ ferromagnet at 
$h=0.05$ calculated by the Green-function theory 
(solid lines), QMC simulations ($\bullet$, $L=16$), and by the method of 
Ref.~\onlinecite{APP07} (dashed lines). In the inset the 
corresponding magnetization is plotted.}
\label{fig_12}
\end{figure}

In two dimensions, the anomaly of $\xi^{zz}$ in the $S=1/2$ ferromagnet is 
more pronounced than in the 1D system and appears already at low fields, 
as can be seen in Fig.~\ref{fig_12}. In contrast to the 1D case, both 
the QMC data and the Green-function theory clearly 
reveal a minimum in addition to the maximum. Note that the statistical 
QMC errors in the interesting temperature region are smaller 
than the size of the symbols. Figure \ref{fig_12} demonstrates the 
qualitative effects of our extended theory 
($\alpha_{2}^{\nu \mu} \neq \alpha_{1}^{\nu \mu}$) on the temperature 
dependence of $\xi^{zz}$ as compared with the simplified approach 
($\alpha_{2}^{\nu \mu} = \alpha_{1}^{\nu \mu}$). Whereas this approach 
yields a slightly better agreement of the magnetization with the QMC 
data (see inset), it fails to describe the minimum-maximum anomaly. 
%
% 
%  fig_13
\begin{figure}[b]
\centering
\includegraphics*[scale=0.581]{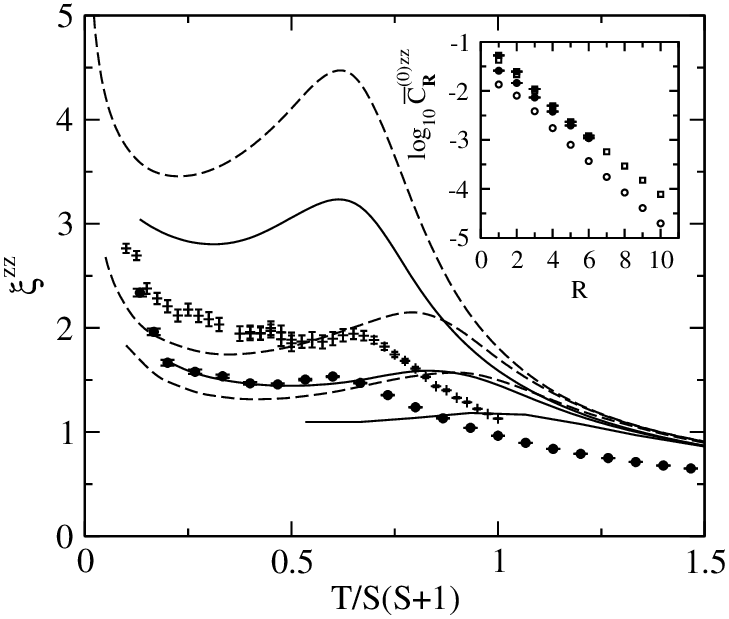}
\caption{Longitudinal correlation length of the 2D ferromagnet with $S=1/2$ 
(solid lines) and $S=1$ (dashed lines) in the fields $h=0.01$, 0.05, and 0.10, 
from top to bottom, as compared with the QMC results at 
$h=0.05$ for $S=1/2$ ($\bullet$, $L=16$) and 
$S=1$ (+, $L=16$). The inset shows the correlation function 
$\bar{C}_{\bm{R}}^{(0)zz}=
\langle S_{0}^{z} S_{\bm{R}}^{z}\rangle - \langle S^{z} \rangle^2$ vs 
$R=|\bm{R}|$ for the 2D $S=1/2$ ferromagnet in the field $h=0.05$ at $T=0.4$ 
and 0.6, from bottom to top, calculated by the Green-function theory 
(open symbols) and by QMC (filled symbols, $L=16$).  
}
\label{fig_13}
\end{figure}

Figure \ref{fig_13} shows the field and spin dependence of the temperature 
behavior of $\xi^{zz}$ in the 2D ferromagnet. As results from the theory, the 
anomaly of $\xi^{zz}$ becomes more pronounced with decreasing field and with 
increasing spin. Let us point out that our QMC data for $h=0.05$ yield a 
minimum and a maximum of $\xi^{zz}$ for both the $S=1/2$ and $S=1$ models 
and give confidence in the results of the theory. As in the 1D model, 
the maximum of $\xi^{zz}$ at $T_m^{\xi}$ is related to the 
maximum of $\bar{C}_{\bm{R}}^{(0)zz}$   by $\lim_{R \to \infty} T_m^{zz}(R)= T_m^{\xi}$
 in all cases shown in Fig.~\ref{fig_13}. The minimum of $\xi^{zz}$ results 
 from the different temperature dependences of $\bar{C}_{\bm{R}}^{(0)zz}$ and 
 $A^{zz}$ in the ansatz (\ref{crz_0}).
In analogy to 
Fig.~\ref{fig_11}, for a more detailed comparison, the inset exhibits the 
correlator $\bar{C}_{\bm{R}}^{(0)zz}$ for $S=1/2$ and $h=0.05$ as function 
of the distance. The relative magnitude of the correlators at $T=0.4$ and 0.6 
may be understood by the maximum in the temperature dependence of 
$\bar{C}_{\bm{R}}^{(0)zz}$.

\subsection{Specific heat}

%  fig_14
\begin{figure}[b]
\centering
\includegraphics*[scale=0.581]{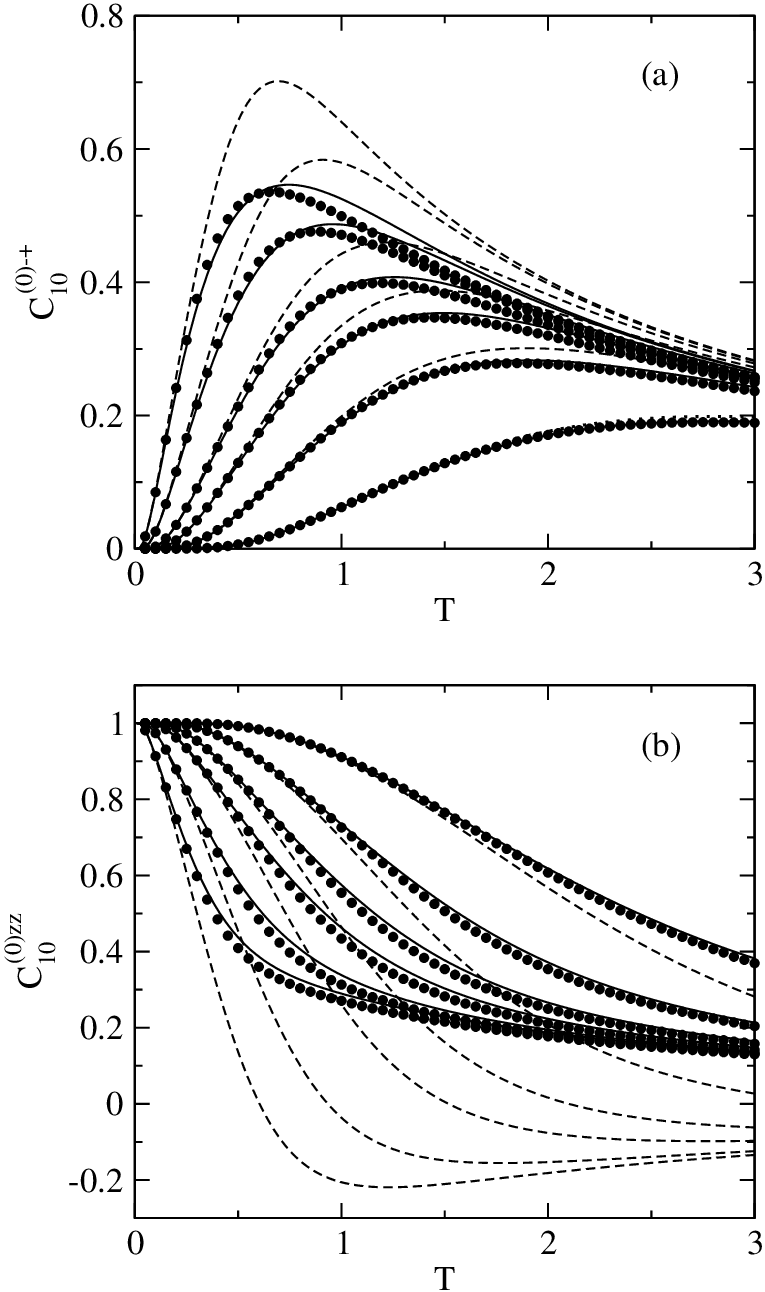}
\caption{(a) Transverse and (b) longitudinal nearest-neighbor two-spin 
correlation functions of the 1D $S=1$ ferromagnet at the fields
$h=0.1, 0.2, 0.4, 0.6, 1.0$, and 2.0, from left to right, obtained by the 
Green-function
theory (solid lines), QMC ($\bullet$, $L=64$), and RPA (dashed lines).}
\label{fig_c10}
\end{figure}
Let us first consider the NN spin correlation functions $C_{10}^{(0)-+}$ 
and $C_{10}^{(0)zz}$ entering the
 internal energy 
$u=-\frac{z}{2} (C_{10}^{(0)-+}+C_{10}^{(0)zz})-h \langle S^z \rangle$. 
As an example, 
for the 1D $S=1$ model they are depicted in Fig.~\ref{fig_c10}, where we obtain  
a very good agreement of the analytical results with the QMC data. 
On the contrary, the RPA results 
for $C_{10}^{(0)-+}$ remarkably exceed the QMC data, and for $C_{10}^{(0)zz}$ 
the RPA yields negative values being incompatible with the ferromagnetic SRO.
%
%  fig_15
\begin{figure}[b]
\centering
\includegraphics*[scale=0.581]{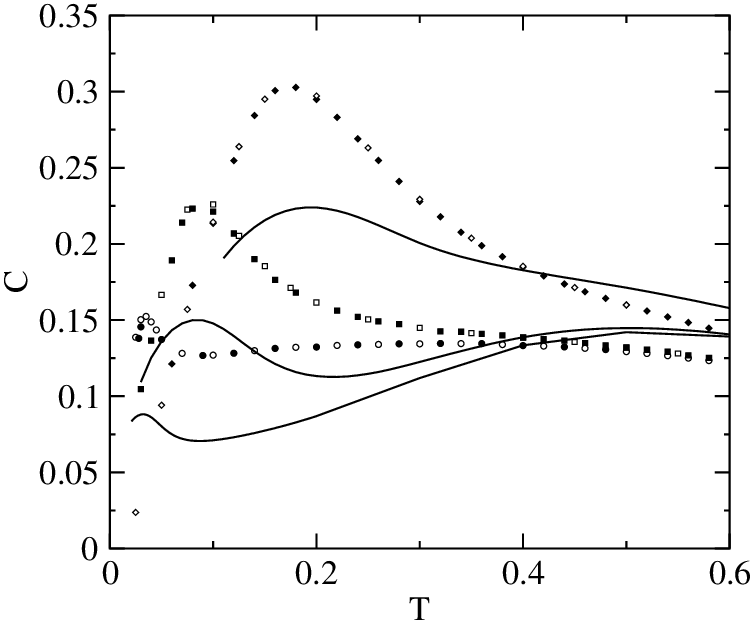}
\caption{Specific heat of the 1D $S=1/2$ ferromagnet obtained by the
Green-function (solid lines) and QMC (filled symbols, $L=128$) methods at low 
fields, 
$h=0.005, 0.03$, and 0.1, from bottom to top, and compared with
the Bethe-ansatz data of Ref.~\onlinecite{JIR04} (open symbols).}
 \label{fig_c_sphalf_1D}
 \end{figure}

In Fig.~\ref{fig_c_sphalf_1D} the specific heat $C=\partial u / \partial T$ for
the 1D $S=1/2$ ferromagnet at low fields is plotted. Again, our QMC data agree 
very well with the Bethe-ansatz results.\cite{JIR04} 
At very low magnetic fields, the 
low-temperature maximum appearing, in the exact approaches at  
$h \lesssim 0.008$, in addition to the high-temperature maximum 
is much better described by the theory than we have found in 
Ref.~\onlinecite{JIR04}. In our Green-function theory this maximum appears  
up to higher fields, $h \leqslant 0.071$, and the deviation of the maximum  
position $T_{m,1}^{C}$ from the Bethe-ansatz and QMC values in the region 
$0.001 \leqslant h \leqslant 0.01$ is less than 8\%. Considering very low 
fields, $h=0.001$ to 0.01 in steps of 0.001, $T_{m,1}^{C}$ 
and the height $C(T_{m,1}^{C})$ are fit by the power laws
\begin{equation}
T_{m,1}^{C} =0.462 \; h^{0.501}, \; \; \; \; C(T_{m,1}^{C})=0.394 \; h^{0.282}.
\label{cpl}
\end{equation}
The exponents are in good agreement with the values of the Bethe-ansatz
results,\cite{JIR04} $T_{m,1}^{C}=0.596 \; h^{0.542}$ and
$C(T_{m,1}^{C})=0.513 \; h^{0.228}$. Note that the specific heat in the
2D model has only one maximum.\cite{JIR04}
%
%
%  fig_16
\begin{figure}[t]
\centering
\includegraphics*[scale=0.581]{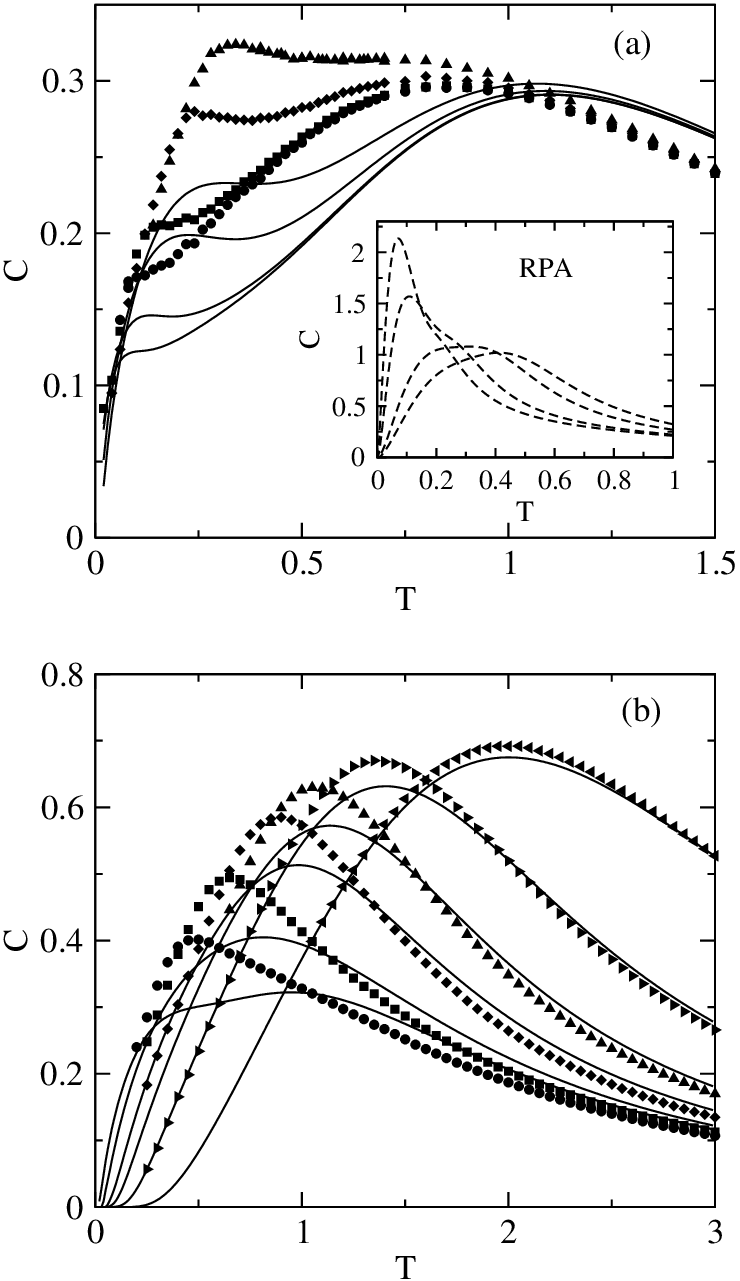}
\caption{Specific heat of the 1D $S=1$ ferromagnet obtained by
the Green-function theory (solid lines) (a) at low fields, 
$h=0.005, 0.01, 0.03$, and
0.05, from bottom to top, with the QMC results for $L=64$ (filled symbols) 
 and (b) at higher fields,
$h=0.1, 0.2, 0.4, 0.6, 1.0$, and 2.0, from left to right, in comparison with
the QMC results for $L=64$ (filled symbols). The inset shows the RPA data at 
the fields given in (a), from top to bottom at $T=0.1$.}
\label{fig_c_sp1_1D}
\end{figure}

\begin{figure}[t]
\centering
\includegraphics*[scale=0.581]{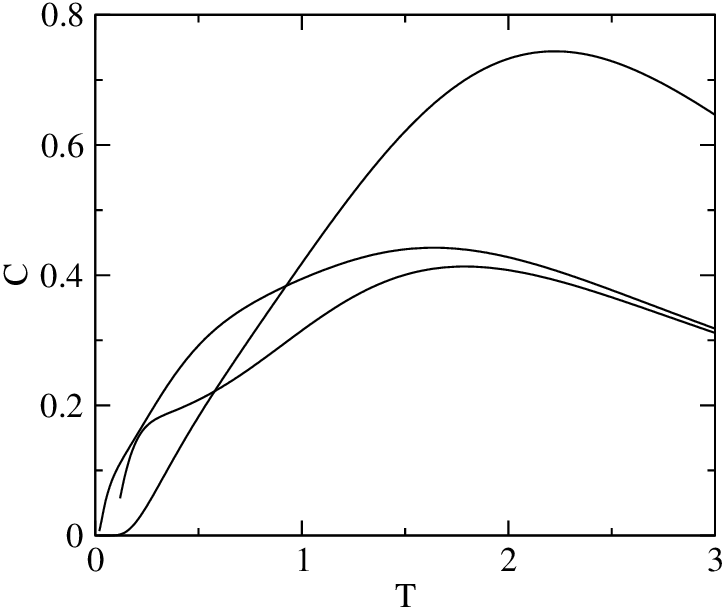}
\caption{Specific heat of the 1D $S=3/2$ ferromagnet calculated by the 
Green-function theory at $h=0.01, 0.1$ and 1.0, at $T=1.5$ from bottom to top.}
\label{fig_c_sp1.5_1D}
\end{figure}

\begin{figure}[b]
\centering
\includegraphics*[scale=0.581]{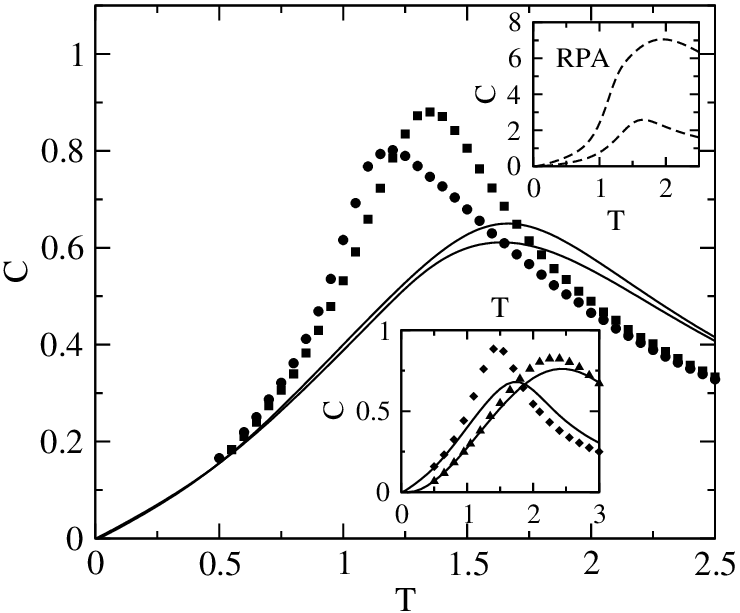}
\caption{Specific heat of the 2D $S=1$ ferromagnet at $h=0.01$ and
0.05, from bottom to top, and, as depicted in the lower inset, at 
$h=0.1$ and 1.0, from left to right, where the Green-function (solid lines) 
and QMC (filled symbols, $L=64$) results   
 are shown. In the upper inset the RPA results for $h=0.01$ and 0.05, from
 top to bottom, are plotted.}
\label{fig_c_sp1_2D}
\end{figure}

Figure \ref{fig_c_sp1_1D} displays the specific heat of the 1D $S=1$ 
ferromagnet. At low magnetic fields, $0.007 \lesssim h \lesssim 0.057$, 
besides the high-temperature maximum, a low-temperature maximum appears 
(see Fig.~\ref{fig_c_sp1_1D}(a)). The position
$T_{m,1}^{C}$ of this maximum obtained by the Green-function theory nearly
agrees with the QMC results. As in the $S=1/2$ case,\cite{JIR04} in RPA 
a double maximum is not obtained (see inset of Fig.~\ref{fig_c_sp1_1D}(a)),
and the values of the specific heat maximum are much higher than the QMC
values which is ascribed
to a poor description of SRO in RPA (see also Fig.~\ref{fig_c10}). The specific 
heat of the 1D $S=3/2$ ferromagnet is shown in Fig.~\ref{fig_c_sp1.5_1D}. 
There is no low-temperature maximum, but only a hump at low enough fields. 
For higher spins qualitatively the same behavior is found. The specific 
heat for the 2D $S=1$ ferromagnet is plotted in Fig.~\ref{fig_c_sp1_2D}. 
As in the case $S=1/2$,\cite{JIR04} in two dimensions only one maximum appears.    
At small fields the position of the maximum in
the Green-function theory is remarkably shifted to higher temperatures 
as compared with the QMC data. Note that the RPA curves at low fields
(see upper inset of Fig.~\ref{fig_c_sp1_2D}) exhibit a too large maximum
height, as was also found in the 1D model (inset of Fig.~\ref{fig_c_sp1_1D}(a)).
%
%  fig_17
% 
%
%  fig_18

From our investigations of the maximum behavior of the specific heat in 
dependence on spin and dimension we conclude that the appearance of two maxima 
is a distinctive effect of quantum fluctuations which decrease with increasing 
spin and dimension. Note that in ferromagnets quantum fluctuations occur 
at nonzero temperatures only, whereas in antiferromagnets they are important 
already at $T=0$. The characterization of the occurrence of two maxima in the 
temperature dependence of the specific heat of the Heisenberg ferromagnet 
as a peculiar quantum effect 
is corroborated by recent QMC simulations of the 1D classical Heisenberg 
model and the 1D $S=1/2$ Ising model in a magnetic field,\cite{W06} where 
only one maximum in the specific heat was found.  

\subsection{Comparison with experiments}

Let us compare our results 
with experiments on $S=1/2$ quasi-1D ferromagnets, where we focus on the 
possible observation of two maxima in the temperature dependence of the 
specific heat as a characteristic feature of 1D ferromagnets in a magnetic 
field. 
% 
% 
% fig_19
\begin{figure}[b]
\centering
\includegraphics*[scale=0.581]{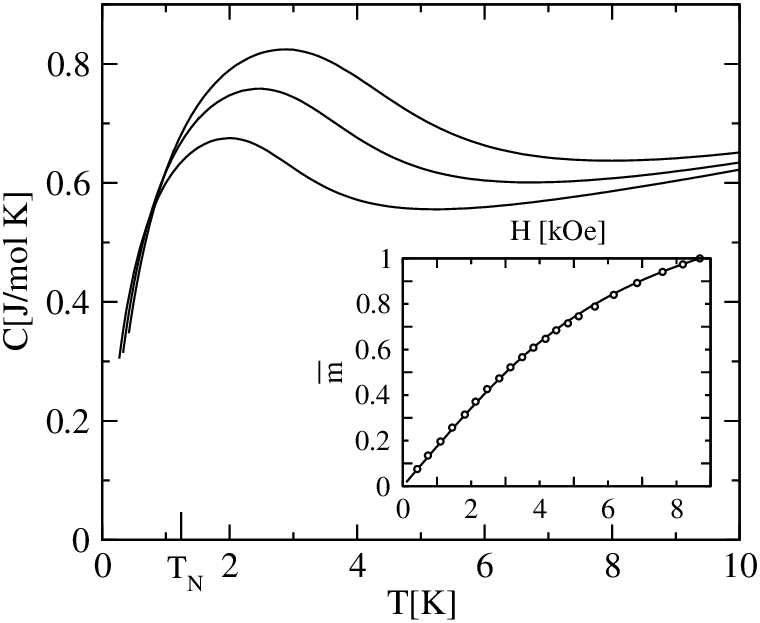}
\caption{Specific heat of the copper salt TMCuC (Refs.~\onlinecite{LW79} and 
\onlinecite{DRS82}, N\'eel temperature 
$T_N=1.24$K), as predicted by the theory for the $S=1/2$ 1D 
ferromagnet in the magnetic fields $H=2$kOe, 3kOe, and 4kOe, from bottom to top, 
with $J=6.18$ meV obtained from the fit of the reduced magnetization 
$\bar{m}= m(H)/m(H=8.7 \text{kOe})$ at $T=4.1$K to experimental data ($\circ$) 
shown in the inset.}
\label{fig_c_exp}
\end{figure}

The copper salt TMCuC [(CH$_3$)$_4$NCuCl$_3$]  
was shown\cite{LW79,DRS82} to be a good 1D Heisenberg ferromagnet which is 
reflected in the small value of the N\'eel temperature $T_N=1.24$K  
for 3D ordering.\cite{DRS82} Determining the exchange energy $J$ by a 
least-squares fit of the theory for $S=1/2$ to the experimental data 
for the magnetization as a function of the 
magnetic field $H$ at $T=4.1$K,\cite{LW79} we obtain $J=6.18$meV and a very 
good agreement with experiments, as can be seen in the inset 
of Fig.~\ref{fig_c_exp}.  
Note that the value of $J$ lies between the values given in 
Ref.~\onlinecite{LW79}   
($J=5.17$meV) and in Ref.~\onlinecite{DRS82} ($J=7.76$meV). 
According 
to the QMC and Bethe-ansatz results for the 1D $S=1/2$ ferromagnet, two maxima 
of the specific heat occur for $h \lesssim 0.008$ or, using the relation 
$h=1.16 \times 10^{-2} H$[kOe]$/J$[meV], for $H \leqslant 4$kOe. 
In Fig.~\ref{fig_c_exp} the specific heat, as predicted by the 
theory using the fit value 
of $J$, is plotted. The low-temperature maximum for $H=2$kOe, 3kOe, and 4kOe   
occurs at $T_{m,1}^{C}=2.0$K, 2.5K, and 2.9K, respectively. The high-temperature 
maximum (not shown in Fig.~\ref{fig_c_exp}) appears at about $T_{m,2}^{C}=37.4$K 
with $C(T_{m,2}^{C})=1.18$J/molK for all fields considered. In the quasi-1D 
system the anomaly of the specific heat at $T_N$, which cannot be described  by 
our theory for a purely 1D system, may mask the low-temperature maximum, if 
$T_{m,1}^{C}$ is not sufficiently larger than $T_N$. At $H=3$kOe (4kOe) we 
have $T_{m,1}^{C}/T_N=2.0$ (2.3). From this we predict that in TMCuC above 
$T_N$ two maxima in the specific heat at moderate magnetic fields, 
$H=3-4$kOe, may be observed. 

Considering the quasi-1D organic ferromagnet p-NPNN 
(C$_{13}$H$_{16}$N$_3$O$_4$) in the $\gamma$ phase with 
$J=0.37$meV,\cite{TTN91,TKI92} where the phase transition at $T_N=0.65$K 
for $H=0$ persists up to $H=1.8$kOe ($T_N \simeq 0.5$K), two maxima of the 
specific heat above $T_N$ cannot be observed, because, at $h \lesssim 0.008$, 
($H \lesssim 0.26$kOe), we have $T_{m,1}^{C} \lesssim 0.19 \text{K} < T_N$. 
The analogous situation, in which the low-temperature maximum in the specific 
heat of the 1D ferromagnet cannot be seen, is found for the following compounds. 
Considering the ferromagnetic chains in the quasi-1D magnet 
$\beta$-BBDTA$\cdot$GaBr$_4$ with $J=0.375$meV,\cite{SGM06} we have 
 $T_{m,1}^{C} \lesssim 0.19$K which is lower than the temperature of the 
 specific-heat cusp, $T_C\geqslant 0.4$K, caused by the interchain coupling. 
 For the CuCl$_2$-TMSO (tetramethylsulfoxide) [DMSO (dimethylsulfoxide)] 
 salts with $J=3.36$ [3.88]meV\cite{SLW79} we get 
 $T_{m,1}^{C} \lesssim 1.7$ [1.96]K being lower than the temperature of the 
 susceptibility maximum, 3.9 [5.4]K, indicating the influence of the 
 antiferromagnetic interchain coupling.

\section{SUMMARY}\label{summ}
In this paper we have developed a second-order Green-function theory 
for the 1D and 2D Heisenberg ferromagnets in a magnetic field 
which extends our previous approach\cite{JIR04} to arbitrary spins and 
by the calculation of the correlation length. 
In addition, we have performed QMC simulations of the $S=1/2$ and $S=1$ models 
on a chain up to $N=1024$ sites and on a square lattice up to $N=64 \times 64$  
using the stochastic series expansion 
method with directed loop updates. The approximate analytical and 
quasi-exact numerical results turned out to be in good agreement, in particular 
for the ferromagnetic quantum spin chains.  
Analyzing the field dependence of the maximum in 
the temperature dependence of the magnetic susceptibility over a much 
broader field region as considered previously\cite{JIR04} we have found 
power laws for the position and height of the susceptibility maximum. 
The transverse and longitudinal correlation 
lengths were shown to have qualitatively different temperature dependences. 
Depending on spin, field, and dimension, the longitudinal correlation length 
$\xi^{zz}$ reveals an unexpected anomaly: with increasing temperature, 
$\xi^{zz}$ exhibits a minimum followed by a maximum. By a detailed
investigation of the specific heat of the Heisenberg chain with arbitrary   
spin, two maxima in its temperature dependence at low magnetic fields  
were detected for $S=1/2$ and $S=1$, whereas for $S > 1$ only 
one maximum appears, as in the 2D case. The existence of two specific-heat 
maxima was identified as a distinctive quantum effect.  
The theory was compared with magnetization 
experiments on the 1D copper salt TMCuC, and predictions for the temperature 
dependence of the specific heat, in particular for the occurrence of two 
maxima, were made which should be measurable experimentally.
\section*{AKNOWLEDGMENTS}
The authors wish to thank J.~Richter, N.~M.~Plakida and S.~Wenzel for 
valuable discussions. This work 
was partially supported (L.~B.~and W.~J.) by the EU through the Marie Curie 
Host Development Fellowship under Grant No. IHP-HPMD-CT-2001-00108 
and the supercomputer time grant No.\ hlz12 of
the John von Neumann Institute for Computing (NIC), Forschungszentrum
J\"ulich.
\appendix*
\section{RANDOM-PHASE APPROXIMATION} \label{app}
It is of interest to compare our results for finite magnetic fields with the 
RPA.\cite{Tjab67} Considering the equation of motion (\ref{eqmtn1})
the Tyablikov decoupling $i \dot{S_{\bm{q}}^{+}} =
\omega_{\bm{q}}S_{\bm{q}}^{+}$ yields
\begin{equation}
\langle\langle
S_{\bm{q}}^{+};
S_{-\bm{q}}^{(n)-}\rangle\rangle_{\omega}=\frac{M^{(n)+-}}{\omega
-\omega_{\bm{q}}}, \;  \; \; \; \omega_{\bm{q}}=z \langle S^z \rangle
(1-\gamma_{\bm{q}})+h,
\label{rpagf}
\end{equation}
with $M^{(n)+-}$ given by Eq.~(\ref{mn}). Comparing the correlation
function $\langle (S_i^z)^n S_i^-S_i^+ \rangle$ resulting from
Eq.~(\ref{rpagf}) with the expression obtained by Eq.~(\ref{id})
multiplied by $(S_i^z)^n$ and using the identity
$\displaystyle{\prod_{m=-S}^{S}}(S_{i}^{z}-m)=0 $, $\langle S^z 
\rangle$ is obtained as\cite{Tjab67}
\begin{eqnarray}
\langle S^z \rangle &=& \{ (S-P)(1+P)^{2S+1}+(1+S+P)P^{2S+1}\} \phantom{mn} 
\nonumber \\
&\times& \{(1+P)^{2S+1}-P^{2S+1}\}^{-1},
\label{rpasz}
\end{eqnarray}
where $P=(1/N) \sum_{\bm{q}}n(\omega_{\bm{q}})$. The transverse two-spin 
correlation functions $C_{\bm{R}}^{(0)-+}$ are calculated from
Eq.~(\ref{rpagf}) for $n=0$ which yields
\begin{equation}
C_{\bm{R}}^{(0)-+}=\frac{2 \langle S^z \rangle}{N} \sum_{\bm{q}} 
n(\omega_{\bm{q}}) \text{e}^{i \bm{q R}}.
\label{rpa_cr}
\end{equation}
The transverse correlation length $\xi^{+-}$ is calculated from the 
long-distance behavior of Eq.~(\ref{rpa_cr}) according to 
Eq.~(\ref{crp_0}). For comparison, the correlation length 
$\xi_{\chi}^{+-}$ may be obtained from the expansion of the static 
spin susceptibility $\chi_{\bm{q}}^{+-}$ 
around $\bm{q}=0$ (cf. Sec.~\ref{sec:res}B). We get 
\begin{equation}
\xi_{\chi}^{+-}=\sqrt{\frac{\langle S^z \rangle}{h}}. 
\label{rpa_ksi}
\end{equation}

The longitudinal correlation functions $C_{\bm{R} \neq 0}^{(0)zz}$ 
cannot be obtained by the RPA, except for the NN correlation function 
$C_{10}^{(0)zz}$ which we evaluate proceeding as in
Ref.~\onlinecite{JIR04} for $S=1/2$. That is, we calculate the internal
energy in RPA starting from the exact representation (\ref{u}) and inserting
the RPA results (\ref{rpagf}) and (\ref{rpasz}),
$C_{10}^{(1)-+}=(1/N)M^{(1)+-}\sum_{\bm{q}} n(\omega_{\bm{q}}) 
\cos q_x$ with $M^{(1)+-}=3 \langle (S^z)^2 \rangle-  
\langle S^z \rangle-S(S+1)$, and
$\langle (S^z)^2 \rangle=S(S+1)-\langle S^z \rangle (1+2 P)$ resulting from
Eq.~(\ref{id}). Moreover, we perform the decoupling
$C_{10}^{(1)zz}=\langle S^z \rangle \langle (S^z)^2 \rangle$. 
From $u, \; C_{10}^{(0)-+}$ and
$\langle S^z \rangle$, the correlator $C_{10}^{(0)zz}$ may be calculated.


\begin{thebibliography}{11}
% [1]
\bibitem{SRF04} \textit{Quantum Magnetism}, Lecture Notes in Physics, 645, 
edited by U.~Schollw\"{o}ck, J.~Richter, D.~J.~J.~Farnell, and R.~F.~Bishop 
(Springer, Berlin, 2004).
% [2]
\bibitem{LW79} C.~P.~Landee and R.~D.~Willett, 
Phys.~Rev.~Lett. \textbf{43}, 463 (1979).
% [3]
\bibitem{DRS82} C.~Dupas, J.~P.~Renard, J.~Seiden, and A.~Cheikh-Rouhou,  
Phys.~Rev.~B \textbf{25}, 3261 (1982).
% [4]
\bibitem{TTN91} M.~Takahashi, P.~Turek, Y.~Nakazawa, M.~Tamura, 
K.~Nozawa, D.~Shiomi, M.~Ishikawa, and M.~Kinoshita, 
Phys.~Rev.~Lett.~\textbf{67}, 746 (1991);
Y.~Nakazawa, M.~Tamura, N.~Shirakawa, \\
D.~Shiomi, M.~Takahashi, M.~Kinoshita, 
and M.~Ishikawa, Phys.~Rev.~B \textbf{46}, 8906 (1992).
% [5]
\bibitem{TKI92} M.~Takahashi, M.~Kinoshita, and M.~Ishikawa, 
J.~Phys.~Soc.~Jpn.~\textbf{61}, 3745 (1992). 
% [6]
\bibitem{SGM06} K.~Shimizu, T.~Gotohda, T.~Matsushita, N.~Wada, 
W.~Fujita, K.~Awaga, Y.~Saiga, and D.~S.~Hirashima, 
Phys.~Rev.~B \textbf{74}, 172413 (2006).
% [7]
\bibitem{SLW79} D.~D.~Swank, C.~P.~Landee, and R.~D.~Willett, 
Phys.~Rev.~B \textbf{20}, 2154 (1979).
% [8]
\bibitem{MTT05} S.~E.~McLain, D.~A.~Tennant, J.~F.~C.~Turner, T.~Barnes, 
M.~R.~Dolgos, Th.~Proffen, B.~C.~Sales, and R.~I.~Bewley, cond-mat/0509194.
% [9]
\bibitem{LPS87} W-H.~Li, C.~H.~Perry, J.~B.~Sokoloff, V.~Wagner, 
M.~E.~Chen, and G.~Shirane, Phys.~Rev.~B \textbf{35}, 1891 (1987);
S.~Feldkemper, W.~Weber, J.~Schulenburg, and J.~Richter, 
\textit{ibid.} \textbf{52}, 313 (1995); \\
H.~Manaka, T.~Koide, T.~Shidara, and I.~Yamada, \textit{ibid.} \textbf{68},
184412 (2003).
% [10]
\bibitem{KV87} G.~Kamieniarz and C.~Vanderzande, Phys.~Rev.~B 
\textbf{35}, R3341 (1987); 
G.~M.~Wysin and A.~R.~Bishop, \textit{ibid.} \textbf{34}, 3377 (1986).
% [11]
\bibitem{FJK00} P.~Fr\"{o}brich, P.~J.~Jensen, and P.~J.~Kuntz,
Eur.~Phys.~J.~B \textbf{13}, 477 (2000);
P. Fr\"{o}brich, P.~J.~Jensen, P.~J.~Kuntz, and A.~Ecker, \textit{ibid.}
\textbf{18}, 579 (2000);
P.~Fr\"{o}brich and P.~J.~Kuntz, \textit{ibid.} \textbf{32}, 445 (2003).
% [12]
\bibitem{HFK02} P.~Henelius, P.~Fr\"{o}brich, P.~J.~Kuntz, C.~Timm, 
and P.~J.~Jensen, Phys.~Rev.~B \textbf{66}, 094407 (2002).
% [13]
\bibitem{SKN05} S.~Schwieger, J.~Kienert, and W.~Nolting, 
Phys.~Rev.~B \textbf{71}, 024428 (2005); 
M.~G.~Pini, P.~Politi, and R.~L.~Stamps, \textit{ibid.} \textbf{72},
014454 (2005).
% [14]
\bibitem{SFM01} R.~Sellmann, H.~Fritzsche, H.~Maletta, V.~Leiner, 
and R.~Siebrecht, Phys.~Rev.~B \textbf{64}, 054418 (2001);
S.~P\"{u}tter, H.~F.~Ding, Y.~T.~Millev, H.~P.~Oepen, and J.~Kirschner, 
\textit{ibid.} \textbf{64}, 092409 (2001).
% [15]
\bibitem{FKS02} P.~Fr\"{o}brich, P.~J.~Kuntz, and M.~Saber, Ann.~Phys.
\textbf{11}, 387 (2002).
% [16]
\bibitem{Tjab67} S.~V.~Tyablikov, in \textit{Methods in the Quantum
Theory of Magnetism} (Plenum Press, New York, 1967).
% [17]
\bibitem{JIR04} I.~Junger, D.~Ihle, J.~Richter, and A.~Kl\"{u}mper, Phys.~Rev.~B
\textbf{70}, 104419 (2004).
% [18]
\bibitem{APP07} T.~N.~Antsygina, M.~I.~Poltavskaya, I.~I.~Poltavsky, and 
K.~A.~Chishko, 
Phys.~Rev.~B \textbf{77}, 024407 (2008).
% [19]
\bibitem{SSI94} F.~Suzuki,
N.~Shibata, and C.~Ishii, J.~Phys.~Soc.~Jpn.~\textbf{63}, 1539 (1994).
% [20]
\bibitem{JIR05} I.~Juh\'asz Junger, D.~Ihle, and J.~Richter, Phys.~Rev.~B
\textbf{72}, 064454 (2005).
% [21]
\bibitem{WI97} S.~Winterfeldt and D.~Ihle, Phys.~Rev.~B
\textbf{56}, 5535 (1997); \textbf{59}, 6010 (1999).
% [22]
\bibitem{KY72} J.~Kondo and K.~Yamaji, Prog.~Theor.~Phys.~\textbf{47},
807 (1972); K.~Yamaji and J.~Kondo, Phys.~Lett.~\textbf{45} A, 317
(1973).
% [23]
\bibitem{ST91} H.~Shimahara and S.~Takada, 
J.~Phys.~Soc.~Jpn.~\textbf{60}, 2394 (1991); \textbf{61},
989 (1992).
% [24]
\bibitem{SRI04} D.~Schmalfu\ss, J.~Richter, and D.~Ihle, Phys.~Rev.~B
\textbf{70}, 184412 (2004); \textbf{72}, 224405 (2005).
% [25]
\bibitem{JA00} P.~J.~Jensen and F.~Aguilera-Granja, Phys.~Lett.~A \textbf{269},
158 (2000).
% [26]
\bibitem{EG79} K.~Elk and W.~Gasser, in \textit{Die Methode 
der Greenschen Funktionen in der Festk\"{o}rperphysik}
(Akademie-Verlag, Berlin, 1979); W.~Nolting, in \textit{Quantentheorie
des Magnetismus}, vol.~2 (B.~G.~Teubner, Stuttgart, 1986).
% [27]
\bibitem{sse1}  A.~W.~Sandvik and J.~Kurkij\"{a}rvi, Phys.~Rev.~B \textbf{43}, 
5950 (1991).
% [28]
\bibitem{sse2}  O.~F.~Syljuasen and A.~W.~Sandvik, Phys.~Rev.~E \textbf{66}, 
 046701 (2002).
% [29]
\bibitem{janke-greifswald06}
W. Janke,
% Monte Carlo Methods in Classical Statistical Physics,\\
in: {\em Computational Many-Particle Physics\/},
% Wilhelm \& Else Heraeus Summerschool, Greifswald, 18--29 September 2006,
edited by H. Fehske, R. Schneider, and A. Wei{\ss}e,
Lect. Notes Phys. {\bf 739} (Springer, Berlin, 2008), pp.~79--140.
% [30]
\bibitem{sse3}  A.~W.~Sandvik, R.~R.~P.~Singh, and D.~K.~Campbell,
Phys.~Rev.~B \textbf{56}, 14510 (1997).
% [31]
\bibitem{Efron}
B. Efron, {\em The Jackknife, the Bootstrap and Other Resampling
 Plans\/}
(Society for Industrial and Applied Mathematics [SIAM], Philadelphia,
 1982).
% [32]
\bibitem{numrec} 
W.~H.~Press, S.~A.~Teukolsky, W.~T.~Vetterling and B.~P.~Flannery,
in \textit{ Numerical Recipes in Fortran 77: The Art of Scientific Computing } 
(Cambridge University Press, Cambridge, 2001).
% [33]
\bibitem{Tak86}  M.~Takahashi,
Prog.~Theor.~Phys.~Suppl.~\textbf{87}, 233 (1986); 
Phys. Rev. Lett. \textbf{58}, 168 (1987).
% [34]
\bibitem{YT86}  M.~Yamada and M.~Takahashi,
J.~Phys.~Soc.~Jpn. \textbf{55}, 2024 (1986).
% [35]
\bibitem{Kop89}  P.~Kopietz,
Phys.~Rev.~B \textbf{40}, 5194 (1989).
% [36]
\bibitem{KC89}  P.~Kopietz and S.~Chakravarty,
Phys.~Rev.~B \textbf{40}, 4858 (1989). 
% [37]
\bibitem{HCP07} A.~Hu, Y.~Chen, and L.~Peng, 
Physica B \textbf{393}, 368 (2007). 
% [38]
\bibitem{Yam90} M.~Yamada, 
J.~Phys.~Soc.~Jpn.~\textbf{59}, 848 (1990).
% [39]
\bibitem{Tak91} M.~Takahashi, 
Phys.~Rev.~B.~\textbf{44}, 12382 (1991).
% [40]
\bibitem{W06} S.~Wenzel, private communication, 2006.
\end{thebibliography}
\end{document}